\documentclass[aps,twocolumn,prb]{revtex4}
\makeatletter
\def\@dotsep{4.5}
\makeatother

\usepackage{graphicx}
\usepackage{amssymb}
\usepackage{natbib}
\usepackage{amsmath}

\newcommand{\comment}[1]{}

\begin{document}

\title{Dynamics of the contact between a ruthenium surface with a single nanoasperity and a flat ruthenium surface: Molecular dynamics simulations}

\author{Alan Barros de Oliveira$^{1,2}$}
\author{Andrea Fortini$^{2,3}$}
\author{Sergey V. Buldyrev$^{2}$}
\author{David Srolovitz$^{2,4}$}
\affiliation{$^{1}$ Departamento de F\'{i}sica, Universidade Federal de Ouro Preto, Ouro Preto, MG 35400-000, Brazil.\\
$^{2}$ Department of Physics, Yeshiva University, 500 West 185th Street, New York, NY 10033, USA.\\
$^{3}$ Theoretische Physik II, Universit\"at Bayreuth, Universit\"atsstra{\ss}e 30 D-95447, Bayreuth, Germany.\\
$^{4}$ Institute of High Performance Computing, 1 Fusionopolis Way, 138632, Singapore.}

\begin{abstract}
We study the dynamics of the contact between a pair of surfaces (with
properties designed to mimic ruthenium) via molecular dynamics
simulations. In particular, we study the contact between a ruthenium
surface with a single nanoasperity and a flat ruthenium surface.  The
results of such simulations suggest that contact behavior is highly
variable. The goal of this study is to investigate the source and
degree of this variability.  We find that during compression, the
behavior of the contact force displacement curves is reproducible,
while during contact separation, the behavior is highly variable.
Examination of the contact surfaces suggest that two separation
mechanism are in operation and give rise to this variability.  One
mechanism corresponds to the formation of a bridge between the two
surfaces that plastically stretches as the surfaces are drawn apart
and eventually separates in shear.  This leads to a morphology after
separation in which there are opposing asperities on the two surfaces.
This plastic separation/bridge formation mechanism leads to a large
work of separation.  The other mechanism is a more brittle-like mode
in which a crack propagates across the base of the asperity (slightly
below the asperity/substrate junction) leading to most of the asperity
on one surface or the other after separation and a slight depression
facing this asperity on the opposing surface.  This failure mode
corresponds to a smaller work of separation. 
those in which a single mechanism operates.  
Furthermore, contacts
made from materials that exhibit predominantly brittle-like behavior
will tend to require lower work of separation than those made from
ductile-like contact materials.

\end{abstract}

\maketitle


\section{Introduction}

Many micro/nanoelectromechanical~\cite{Rebeiz:2003} systems
(MEMS/NEMS) are based upon mechanical contacts that are only a few micrometers large.
MEMS electrical switches must maintain high conductivity and mechanical 
reliability operating at radio-frequencies during their lifetime which may extend to years. 
However, MEMS switches have been hindered by a lack of reliability. 
Extensive surface damage leads to failure of some MEMs devices
after only several million open/close cycles.~\cite{Chen:2007fj} The resulting
surface damage has been studied experimentally with atomic
force microscopes in order to understand the effect of adhesion,
thermal dissipation, and contamination.~\cite{Hyman:1999,Erts:2002,TORMOEN:2004}

Although a variety of approaches have been used to study ideally flat surfaces~
\cite{Kennedy:1974,Komvopoulos:1988,Kral:1996,Mesarovic:1999,Komvopoulos:2001} the contact surfaces of MEMS are rough~\cite{Rezvanian:2007} with  a high density of nanoscale asperities. 
Probabilistic,~\cite{Greenwood:1966,Larsson:1999} and fractal~\cite{Berry:1980,Borodich:1999,Majumdar:1990,Majumdar:2001} approaches 
have been used to describe the correct topography of metal surfaces. 
MEMS contacts are intrinsically multiscale, involving atomic bonding, defect, and fracture nucleation at the subnanometer scale, 
plastic and elastic deformation at the scale of a single nanoasperity as well as elastic deformation at the level of the entire 
MEMS switch. While the macroscopic elastic deformation in the presence of adhesion can be successfully characterized in terms of 
Johnson, Kendall and Roberts (JKR) theory\cite{KL:1971} and its more recent extensions, \cite{Maugis:1995,Mittal:2001,Lin:2008} 
the plastic deformation and adhesion at the level of
nanoasperity is difficult to treat analytically yet can be effectively modeled
using molecular dynamic simulations. In this sense, a multiscale approach in which the microscopic elastic deformation of the contact 
is modeled by classical finite element simulations while the nanoscale contacts at the single
asperity level are replaced by inelastic springs, with parameters taken from molecular dynamic simulations, 
is a promising avenue of MEMS research. 

A fundamental difference between mechanical behaviors of microscopic MEMS
contacts and single, nanoscale asperity contacts is that the latter may be governed by random events such as defect nucleation
and fracture, while in the former these random events are likely to be averaged out due to the large number of asperities 
forming a macroscopic contact. Nevertheless, under some conditions, randomness at the nanoscale may lead to catastrophic 
failure of the switch such as its permanent stiction.  Thus, it is important to characterize the reproducibility and variability of the mechanical
behavior at the single asperity level. 

In this paper, we focus on the statistical properties of the formation and breaking of the ruthenium single asperity contacts.
Ruthenium contacts have proven to be more reliable than gold contacts, routinely surviving millions of cycles without significant degradation. 
The behavior of Ru single asperity contacts has recently been studied by molecular dynamic simulations and compared to 
that of gold contacts~\cite{Fortini:2008}. In that work, a new embedded atom potential~\cite{Daw:1983fk,Daw:1984uq} was 
developed which accurately reproduces several key structural, thermodynamic and mechanical properties of ruthenium, including its hexagonal close 
packed lattice structure, elastic constants, work of adhesion, and stacking fault energy. It has been shown that while the behavior of Ru and Au 
asperity during compression is qualitatively similar, their behavior during separation is distinctively different. 

Au nanoasperity contacts~\cite{Cha2004,Song2006,Song2007,Song2007a,Sorensen:1998,Kuipers:1993,Kizuka:1998} demonstrate typical ductile behavior. Namely, the asperity after sticking to an opposite substrate during compression, forms a symmetric bridge, which gradually elongates on separation, while the neck of this bridge is getting thinner until its diameter becomes comparable to a single atom size. 
At this moment the contact breaks. In this case plastic deformation occurs continuously without formation of a crack. This behavior is typical of ductile materials. 
On the other hand, a Ru nanoasperity contact was reported~\cite{Fortini:2008}  to  show a behavior which is more typical for brittle materials like glasses. Namely, the contact separation is characterized by crack formation and a sharp drop-to-zero of the tensile force during the unloading stage. 

In the present study we find that the behavior of Ru nanoasperity contacts is intrinsically chaotic. Depending on a slight change of initial conditions due to thermal noise, Ru bridges may break in brittle-like manner with a sharp drop of the tensile force or in a ductile-like manner which resembles the behavior of Au contacts.   

The brittle and ductile failure mechanisms are macroscopic phenomena,\cite{Rice:1974} while brittle materials like glass shatter very quickly, ductile materials like metals, can be deformed continuously. The origin of the mechanism is atomic in nature~\cite{Bue2008}. At the atomic level the crystalline structure and the corresponding available glide planes for the motion of dislocations are the most important factor in the distinction between brittle and ductile materials.
A non crystalline material, like glass, is brittle because dislocation movement is not possible. A crystalline material always has a certain degree of plasticity and the competition between a brittle and ductile behavior depends on how the interatomic bonds close to the tip of a crack respond to the locally large interatomic forces. Therefore at a microscopic level a material is more ductile or brittle depending on its crystalline structure and interatomic forces. A good operative definition of brittle and ductile behavior can be given by the separation mode. In Ref.~\cite{Fortini:2008},  it was hypothesized that Ru contacts are more brittle than Au contacts due to the fact that the HCP lattice of Ru has fewer slip systems than the FCC lattice of Au and also due to the fact that Ru stacking fault energy (which is activated in plastic deformations) is 16 times larger than that of Au, while the Ru surface energy (which is activated during fracture formation) is only 3 times larger than that of Au. Our present study which shows the variability of the behavior of Ru contacts on separation does not contradict this hypothesis.

\section{Methods}
\label{sec:met}
We perform the molecular dynamics simulations using the LAMMPS package.
\cite{Plimpton:1995om} Pressure and temperature were kept constant
using the Nos\'e-Hoover thermostat \cite{Hoover:1985yq} and barostat
\cite{Hoover:1986vn} with a time step of $\delta t=0.0025$ ps. 

In the contact simulations, we create two substrate slabs facing one
another. The substrate surfaces are flat and parallel to the $x-y$ plane and consist of 18 (0001) atomic planes of Ru hexagonal close packed (hcp) lattice.  Upon the lower substrate we place a homoepitaxial asperity, as shown in Fig.~\ref{fig:brittle-like-ductile-like}. 
The atoms of the top two atomic layers of the upper substrate and the bottom two layers of the lower substrate are not updated according to Newton's equation of motion (as normal in molecular dynamics) but are displaced in the $z$-direction at constant rates in order to bring the surfaces into/out of contact and to compress/separate the substrates in $z$-direction.  The $x$- and $y$-coordinates of the atoms in these layers are scaled to maintain zero net stress in the $x$- and $y$-directions.  The system is periodic in the $x$- and $y$-directions.  These boundary conditions are imposed to simulate large, finite substrates. 

The simulation cell contains a total of 62150 atoms of ruthenium, whose
dimensions are $L_x=97.4$, $L_y=103.1$, and $L_z=121.7 \AA$ . 
The asperity is constructed as a homoepitaxial cubic island with dimensions 32.48, 37.50, and 27.86 \AA 
~(2891 atoms). 
Prior to the contact simulation the system is annealed in four stages,
each one taking $100ps$
to be completed. First, the temperature is increased linearly from
$T=3K$ to $T=1650K$, close to the melting temperature of the ruthenium
model used in this work, and subsequently reduced linearly to $T=3K$. 
In the third stage, the system is heated again until it achieves $T=600K$. Finally,
in the fourth stage, it is equilibrated at $T=300K$. 

After the asperity annealing is complete, we randomly assign atomic velocities from a 
Maxwell distribution at temperature $T=300K$.
The randomization of velocities is done in order to examine the variability of the contact simulation results.  Next, we
displace the upper substrate toward the lower substrate at 0.07$\textrm{\AA}/ps$,
while holding the lower substrate fixed (i.e., we hold the two atomic
layers at the bottom of the lower substrate fixed). When the
distance between the top of the upper substrate and the bottom
of the lower substrate reaches a predetermined limit, the sign of the velocity of the
upper substrate is reversed. The simulation continues until the upper
and lower substrates are completely separated. The $z$-component
of the force on the upper substrate is calculated during the simulation
in order to determine the force-displacement relation. 

The embedded atom method (EAM) potential used to describe the interaction between Ru atoms may be written in the form~\cite{Finnis:1984ca} 
\begin{equation}
U=\sum_{i=1}^{N-1}\sum_{j=i+1}^{N}V(r_{ij})+\sum_{i=1}^{N}F\left(\rho_{i}\right),\label{eq:eam}
\end{equation}
where the subscripts $i$ and $j$ indicate each of the
$N$ atoms in the system, $r_{ij}$ is the distance between atoms
$i$ and $j$, $V(r_{ij})$ is a pairwise potential, $F(\rho_{i})$
is the embedding energy function, $\rho_{i}=\sum_{j}\Phi(r_{ij})$
models the electron density at the position of atom
$i$, and $\Phi(r_{ij})$ is another pairwise potential. The functional forms and parameters 
for $V(r_{ij})$,  $F(\rho_{i})$, and $\Phi(r_{ij})$ can be found in Ref. ~\cite{Fortini:2008}.
The interatomic potential adopted for the Ru atoms
in this work reproduces the elastic constants and cohesive energy of Ru, 
and also gives stacking fault energy and surface energies that are in reasonable agreement with the experiment 
(for a comparison between experiment and model for several quantities of interest
see  Ref. ~\cite{Fortini:2008}). 
Stacking fault energies play an important role in plastic deformation and surface energy determines the work of adhesion; 
both are key elements in contact formation and separation phenomena. 
Despite the melting temperature of this model is low (1792 K)  compared to the experiments (2607 K),
this is not a problem since all simulations were performed at temperatures less than a quarter of the melting temperature of the model.

\section{Contacts}
\label{contacts}

\begin{figure*}
\includegraphics[scale=0.13]{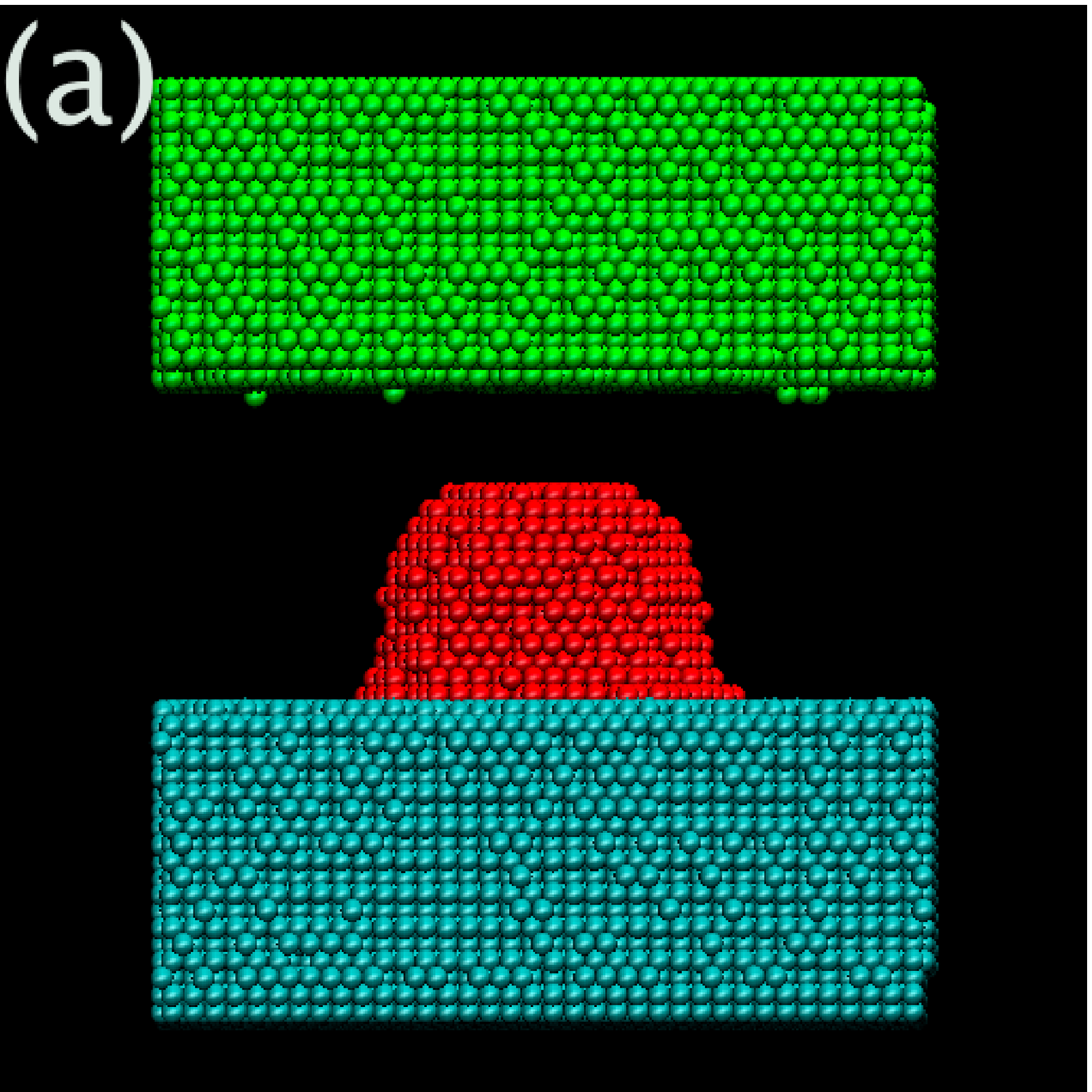}\includegraphics[scale=0.13]{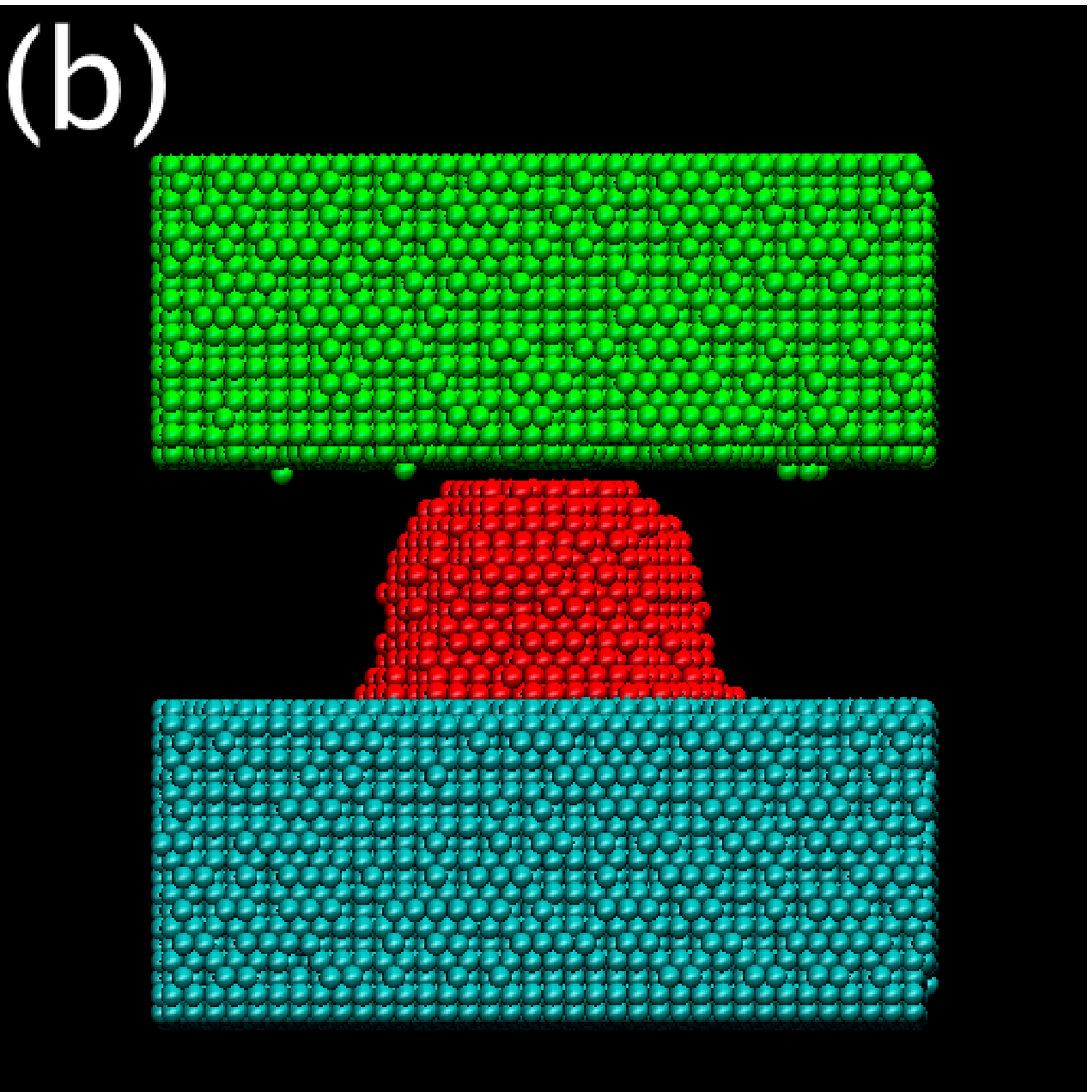}\includegraphics[scale=0.13]{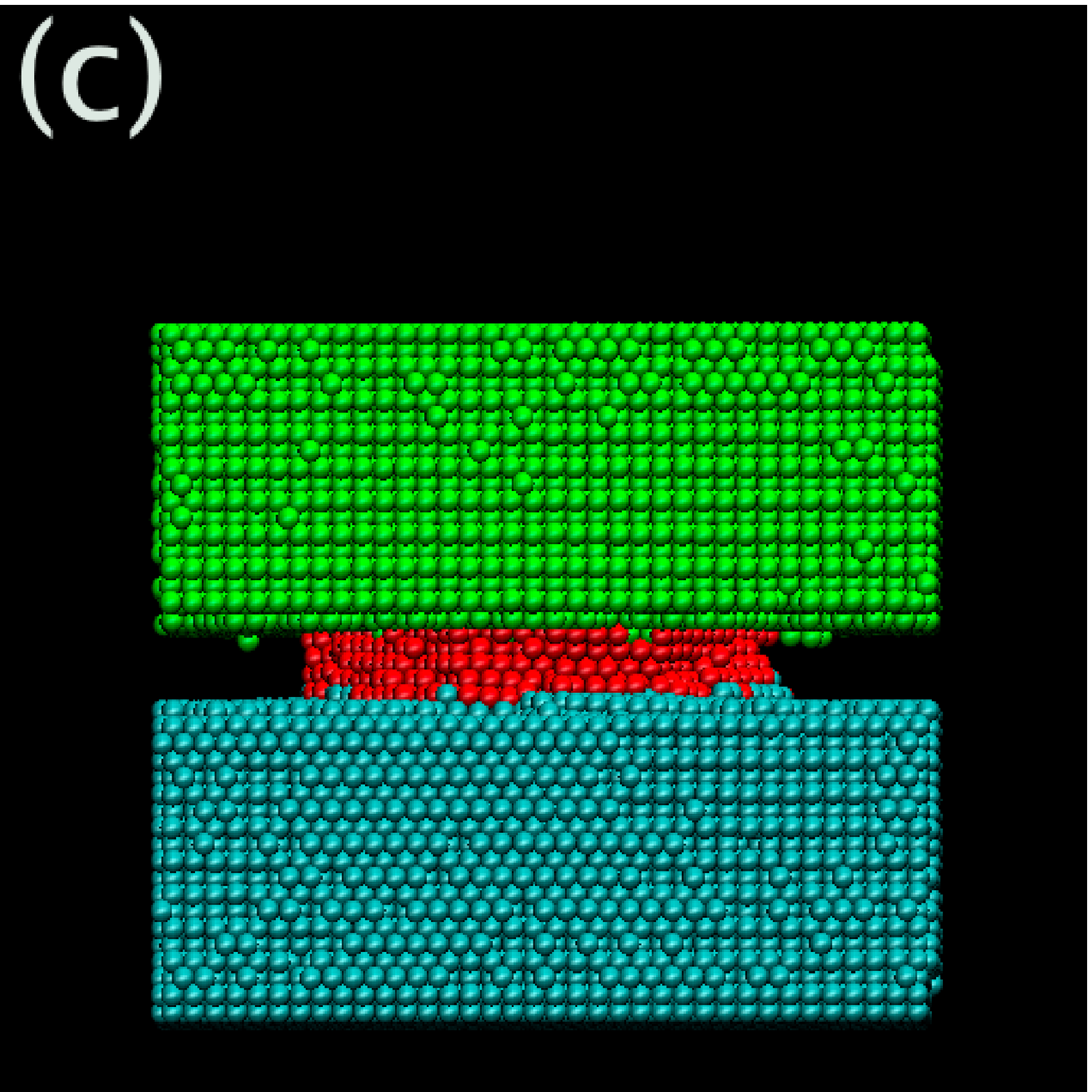}\includegraphics[scale=0.13]{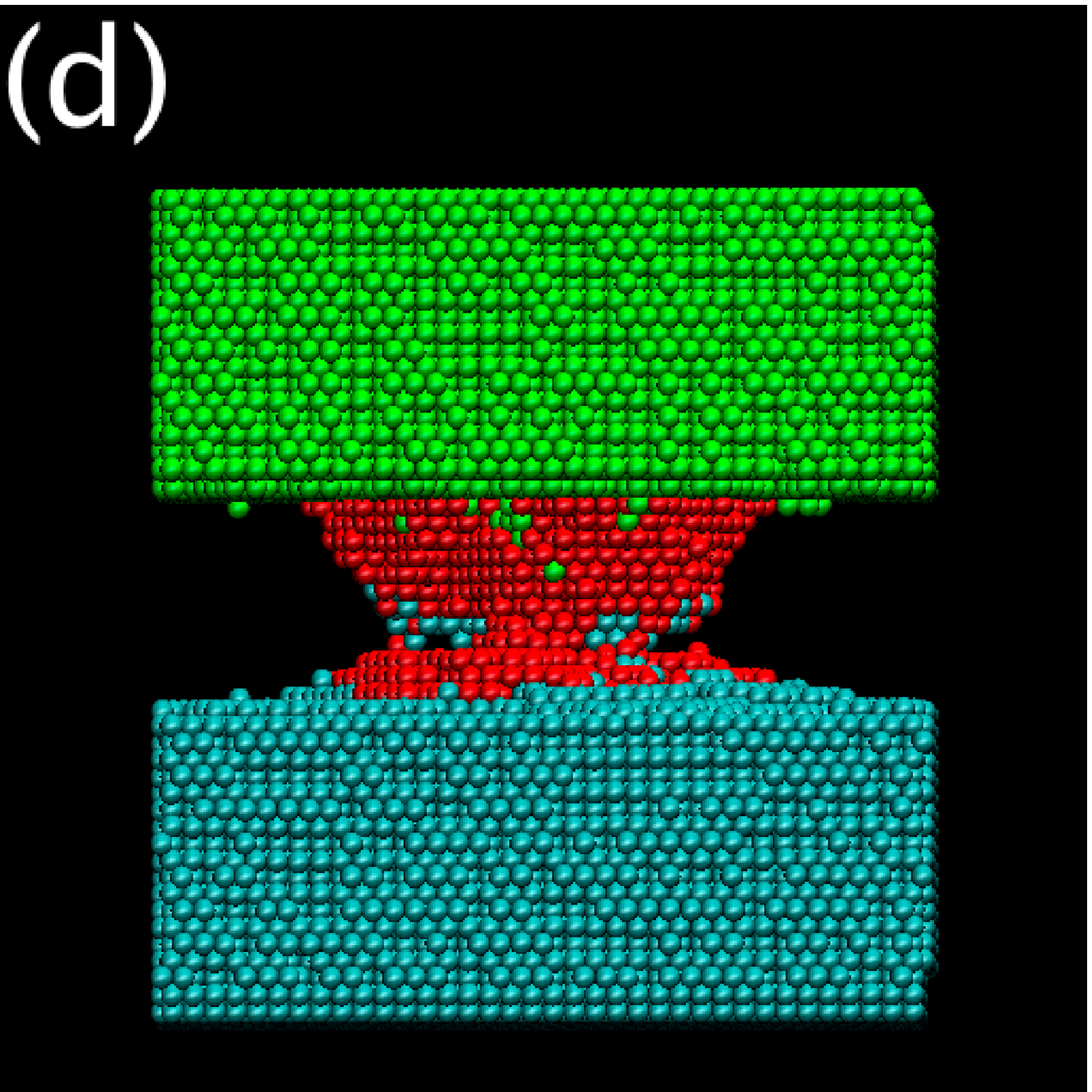}\includegraphics[scale=0.13]{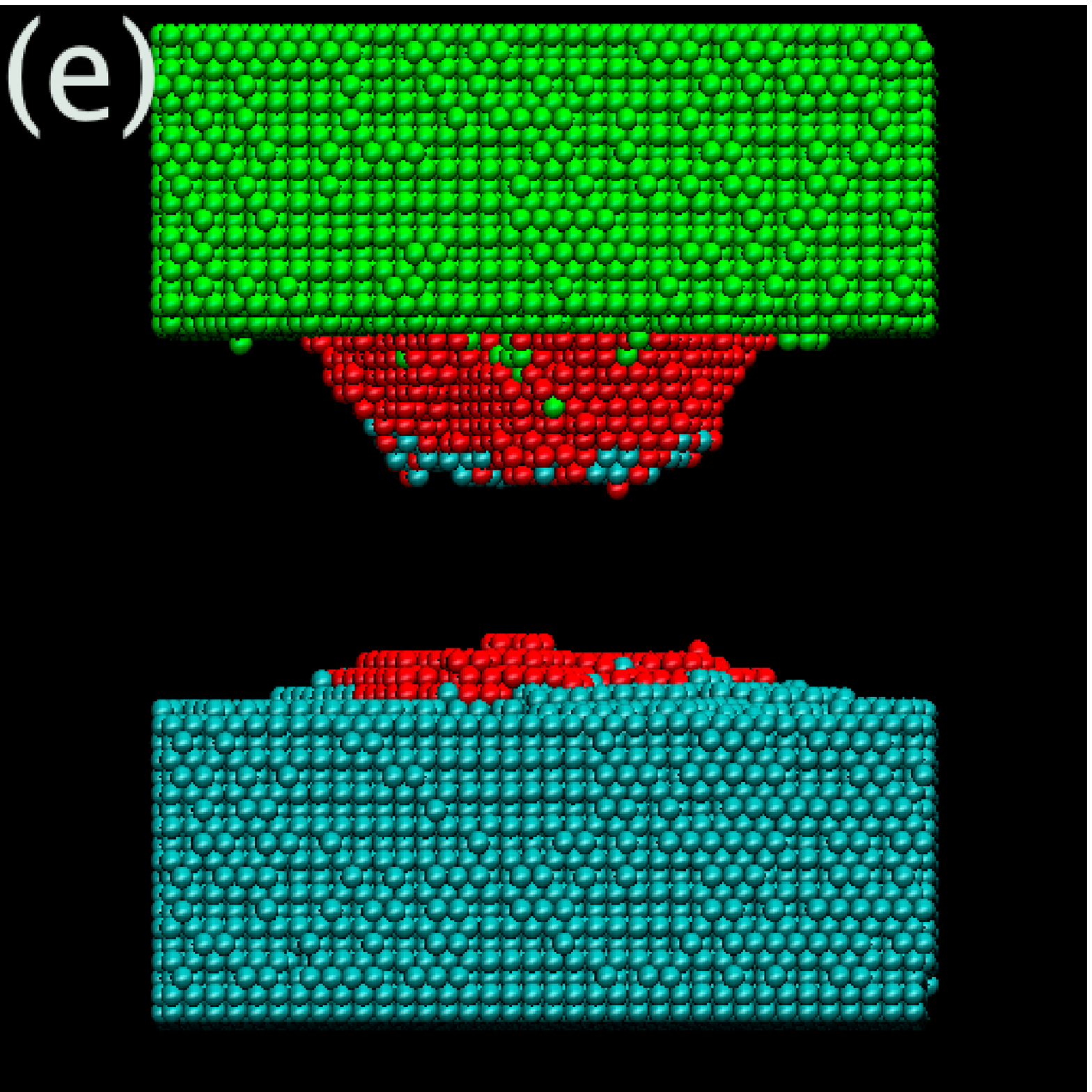}

\includegraphics[scale=0.13]{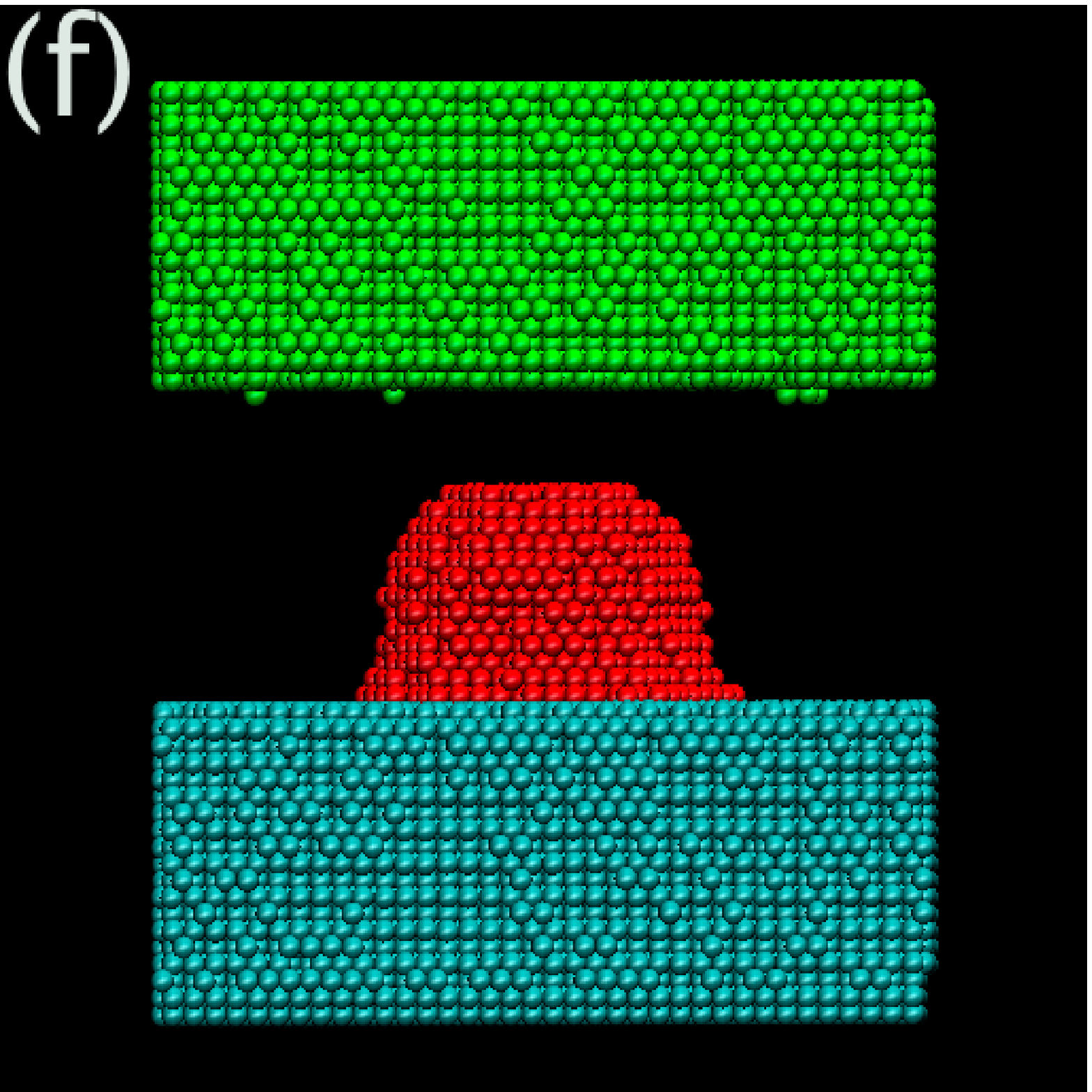}\includegraphics[scale=0.13]{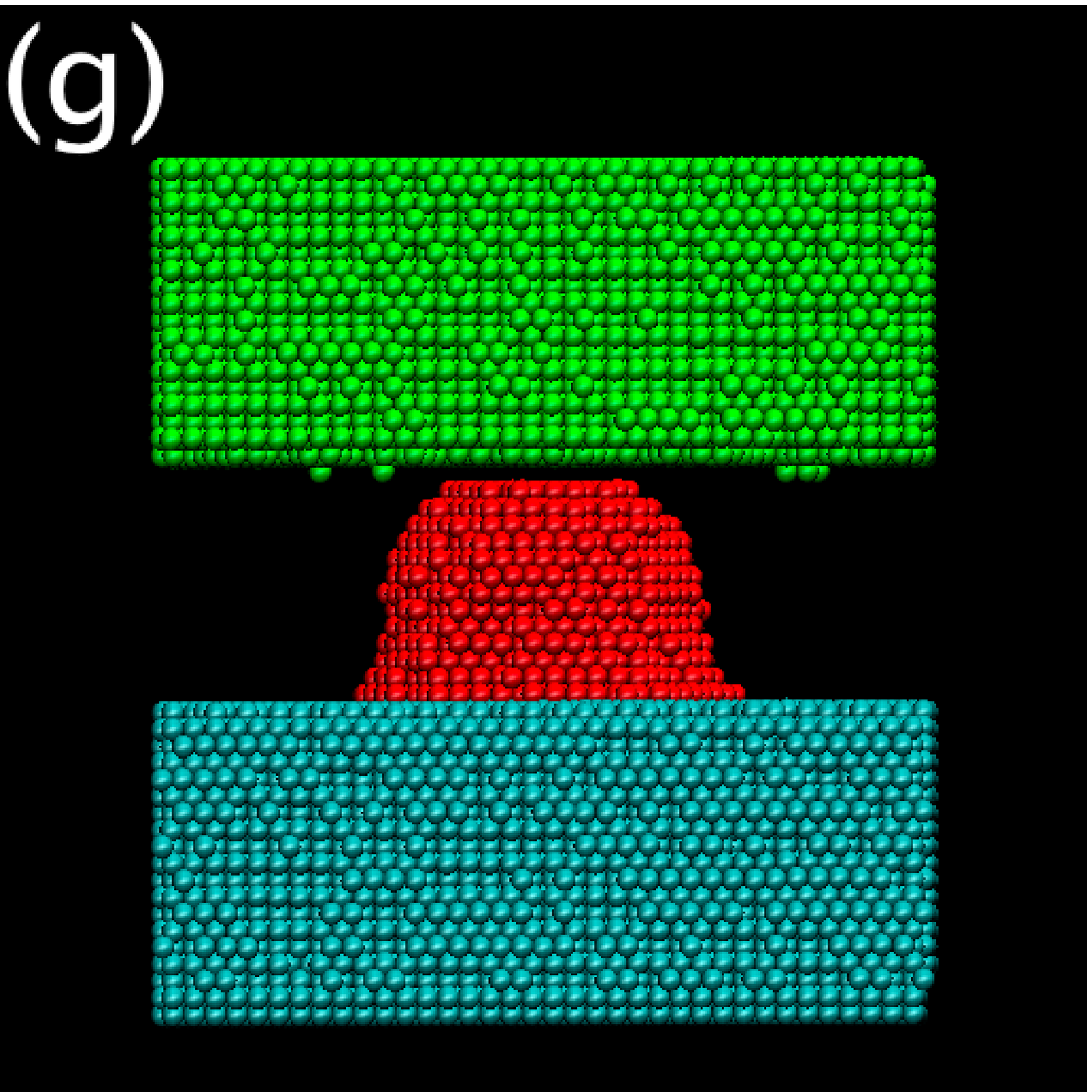}\includegraphics[scale=0.13]{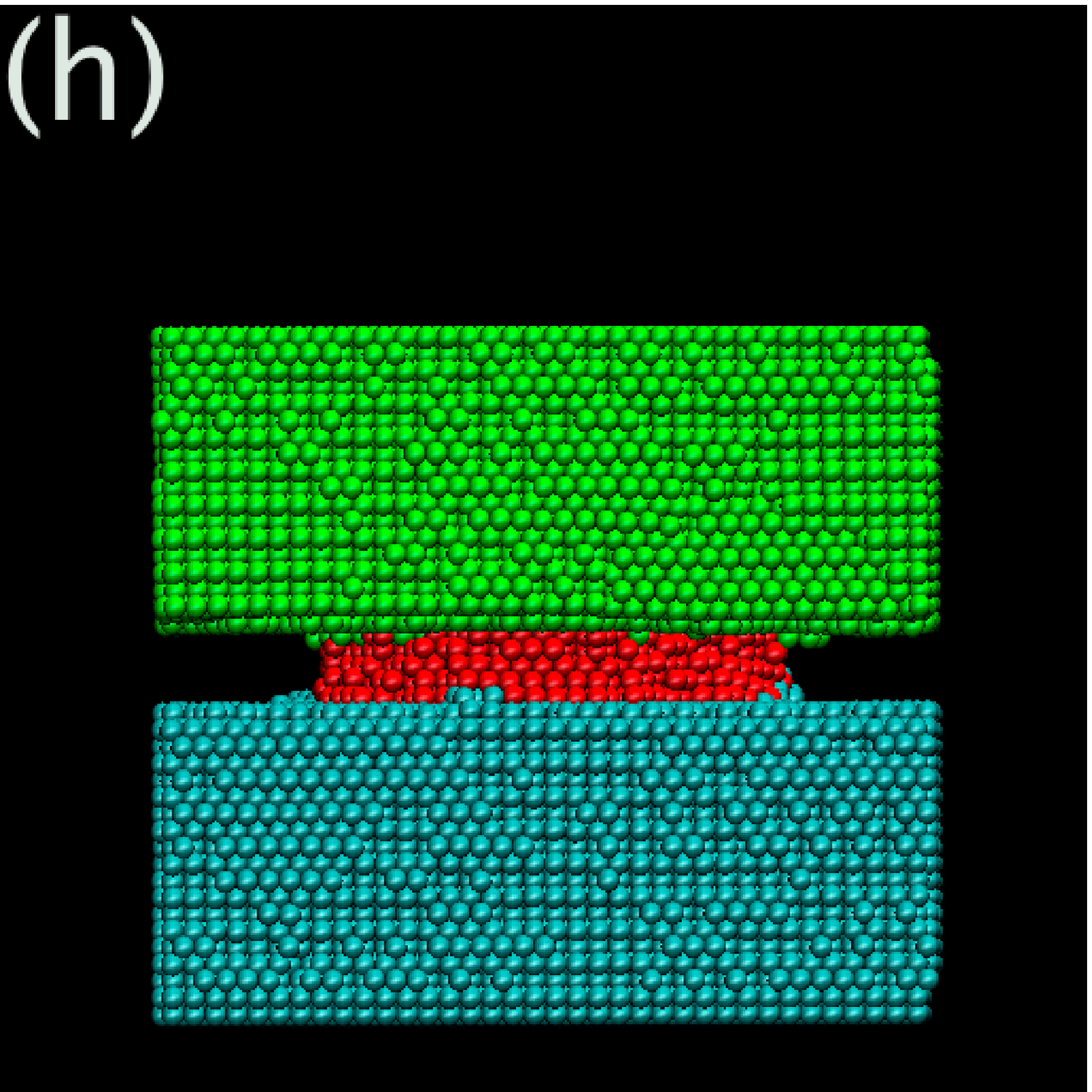}\includegraphics[scale=0.13]{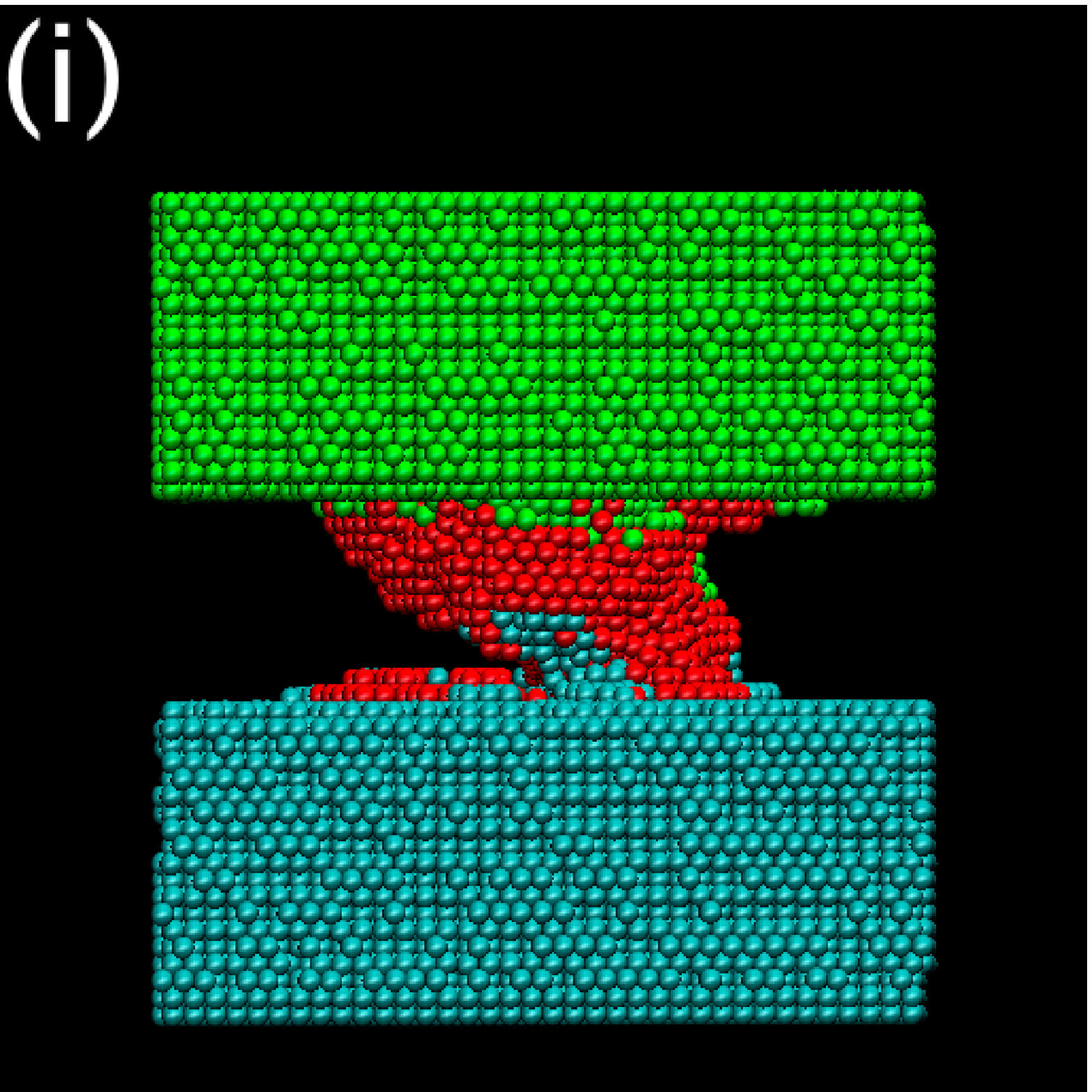}\includegraphics[scale=0.13]{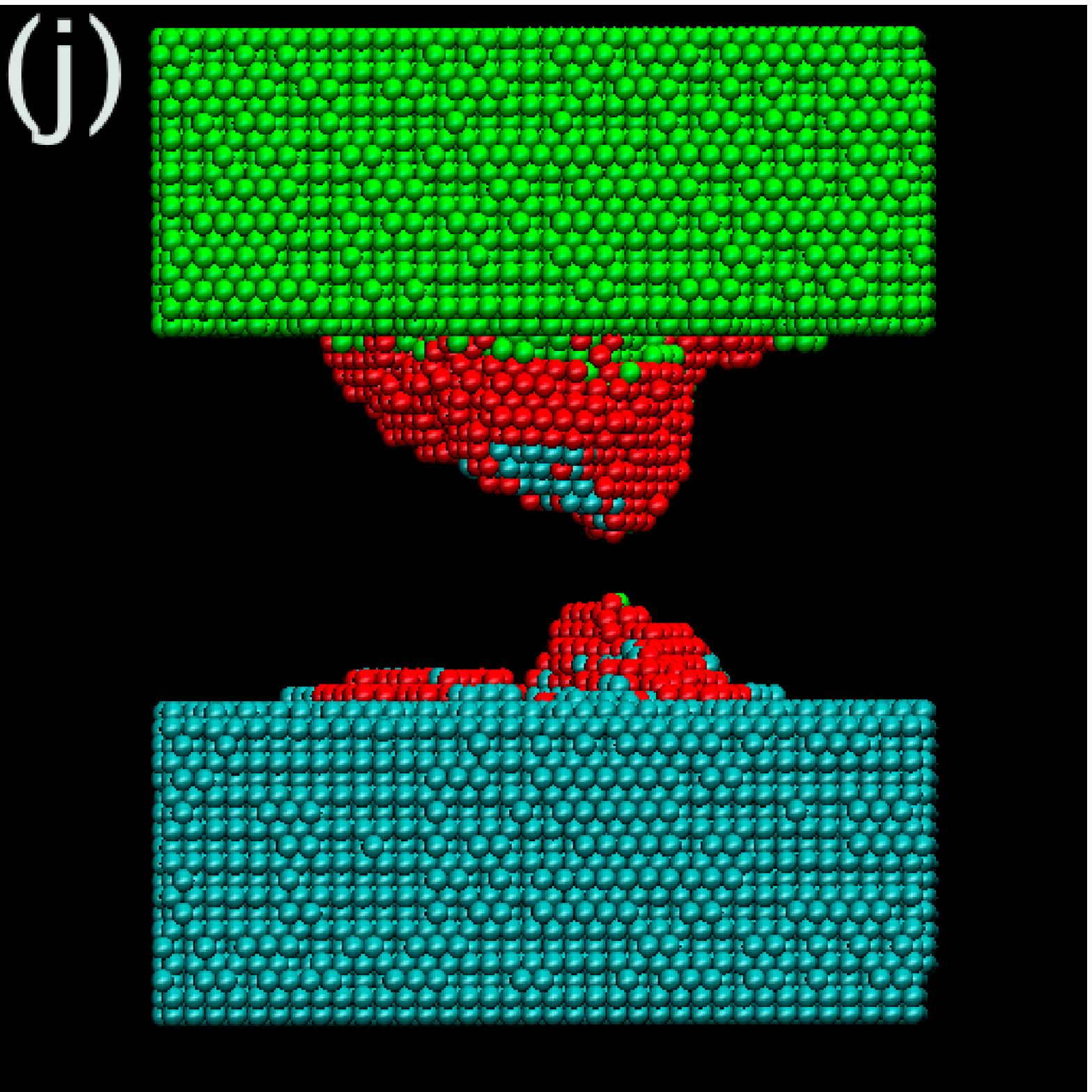}

\caption{The atomic configuration of the system for different displacements of the upper substrate (see also the following figure) as the two surfaces are brought together (loading) and separated (unloading). 
Panels (a)-(e) and (f)-(j) show the evolution of the system for two different distributions of the initial atomic velocities (see the text for details).  The maximum displacement of the upper substrate is $z_{m} = 21.23\textrm{\AA}$.
Panels (a)-(e) show a case in which the asperity behaves in a brittle-like manner, while panels (f)-(j) suggest ductile-like behavior.
\label{fig:brittle-like-ductile-like}}
\end{figure*}

\begin{figure}
\includegraphics[clip,scale=0.32]{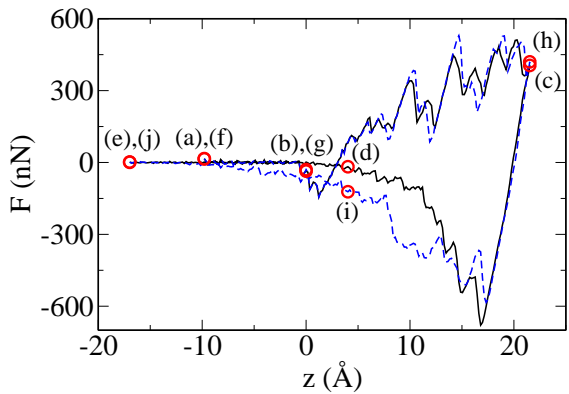}

\caption{Force versus displacement for $z_{m}=21.53\textrm{\AA}.$ A positive force corresponds to pushing the surfaces together; negative to pulling them apart.  The zero of the displacement is set at the separation between surfaces corresponding to first interaction as the surfaces are brought into contact.  The continuous line corresponds to brittle-like behavior, as represented by panels  (a)-(e) of Fig. \ref{fig:brittle-like-ductile-like}, and the dashed line corresponds to ductile-like behavior, as represented by panels (f)-(j) of the same figure.  The points labeled as (a)-(j) here, correspond to the configurations shown in Fig.~\ref{fig:brittle-like-ductile-like} denoted by the same letters.
\label{fig:Force-versus-displacement}}

\end{figure}

Figure \ref{fig:brittle-like-ductile-like} shows a typical evolution of the Ru system during a single 
nanoasperity contact simulations.
In this figure, we show two complete simulations corresponding to two different
realizations of the initial atomic velocities (exactly the same initial contact geometry). 
We can see that, depending on the
initial conditions, the plastic behavior of the contacts during separation can be either ductile-like or brittle-like. 
In both cases in Fig. \ref{fig:brittle-like-ductile-like}, the system was subjected to the same degree of compression, 
and the upper and lower panels correspond to the same stage in the loading/unloading cycle.  
Figure \ref{fig:Force-versus-displacement} shows the force in the $z$-direction 
as a function of the displacement of the upper substrate for the two cases seen 
in Fig.\ref{fig:brittle-like-ductile-like} [points (a)-(j) 
in this figure refer to panels (a)-(j) in Fig. \ref{fig:brittle-like-ductile-like}].

The upper panels in Fig. \ref{fig:brittle-like-ductile-like}, (a)-(e), show a case in which a crack forms near the base of the asperity and propagates 
across the asperity (nearly horizontally) as the two substrates are pulled apart
{[}see Fig. \ref{fig:brittle-like-ductile-like}(e)]. We characterize this behavior as
brittle-like.  The lower panels in Fig. \ref{fig:brittle-like-ductile-like}, (f)-(j), show a nominally identical compression/separation simulation, but with different, randomly chosen, initial atomic velocities. 
In this case, the behavior can be described as ductile-like in the sense that there is significant asperity necking (thinning) prior to separation. We can see the difference between the brittle-like and ductile-like scenarios 
by comparing panels (d) and (i), and (e) and (j) in Fig. \ref{fig:brittle-like-ductile-like}.
We see that in (d), the lower substrate has almost no contact with
the asperity anymore and the tensile force drops to zero (see Fig.~\ref{fig:Force-versus-displacement}). In contrast, in (i) the asperity forms a bridge between the two substrates. Gradual elongation of this bridge corresponds to the slow decrease of the tensile force between the two substrates (see Fig.~\ref{fig:Force-versus-displacement}). In this case, the tensile force persists to much larger separations than in the upper panel. Panels (e) and (j) show very different final configurations for the two cases. 
While in (e) we observe a horizontal crack at the base of the asperity, panel (j) presents 
 two opposing asperities, which we call ``stalactite''  and ``stalagmite''.  Stalactite/stalagmite-like
final structures are commonly seen in ductile-like materials such as gold. \cite{Fortini:2008}

In Fig.~\ref{fig:Force-versus-displacement}, 
zero displacement corresponds to the first contact between
the top substrate and the asperity {[}see Fig. \ref{fig:brittle-like-ductile-like}(b)
and (g)]. At large separation, which corresponds to negative displacement {[}see
Fig. \ref{fig:brittle-like-ductile-like} (a) and (f)], the force is zero (see
Fig. \ref{fig:Force-versus-displacement}). As the two substrates
approach each other, the force becomes negative because of the short
range attractive interactions between the atoms on opposing substrate
surfaces (this distance is defined by the interaction range of the interatomic
potentials). This attraction elastically stretches the two materials
and there is a jump-to-contact \cite{Smith:1989,Landman:1990,Untiedt:2007}
indicated by the first negative spike in the force-displacement curve (e.g., note in Fig. \ref{fig:Force-versus-displacement} that
the force becomes negative at $z=0$). As the displacement increases
beyond this point, the system starts to compress. There is an
approximately linear rise in the force with displacement, punctuated
by a series of relatively sharp drops. The linear increase between
the drops corresponds to elastic compression. The sharp drops correspond
to defect generation, migration and/or annihilation events. When the sign of the
substrate velocity changes {[}panels (c) and (h) in Fig. \ref{fig:brittle-like-ductile-like}],
the system begins to recover. The unloading
is initially characterized by a long linear elastic region.
Eventually, the force reaches a minimum (i.e., a maximum tensile force),
following which the tensile force slowly decreases to zero over a
long displacement range. In this region, the overall force-displacement
trend is also interrupted by sharp jumps. During the ductile-like run (f)-(j) the two substrates separate
at larger (negative) displacement than that at which the initial contact
occurred (i.e., zero displacement). On the other hand, in the brittle-like run (a-e), separation occurs at nearly the same displacement as where the original contact occurred.

\begin{figure}
\includegraphics[clip,scale=0.32]{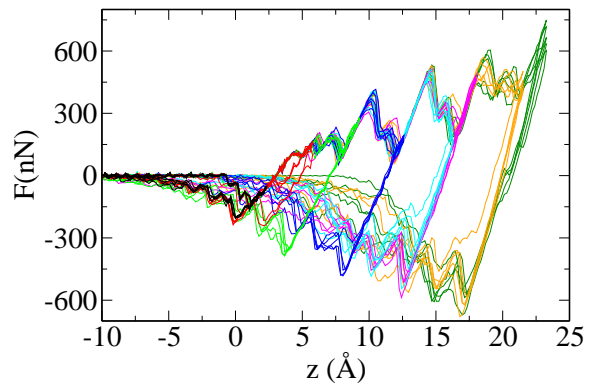}

\caption{Force versus displacement of the upper substrate towards the
lower one for different maximum displacements, $z_m$. The curves show little variation on loading.  We show five independent runs for each of eight maximum displacements: $z_{m}=2.23,$ $5.78,$ $9.28,$ $12.6,$ $16.28,$ $18.03,$
$21.53,$ and $23.28$ \AA. Each run begins from the same atomic configuration but with atomic velocities chosen at random from the same (Maxwell) distribution.
\label{fig:Spaghetti-curve}}

\end{figure}

Figure \ref{fig:Spaghetti-curve} shows the force versus displacement
for different maximum displacements $z_{m}$ (i.e., the displacement at which the sign of the substrate velocity switches). 
The cases analyzed in this work are $z_{m}=2.23,$ $5.78,$ $9.28,$ $12.6,$ $16.28,$ $18.03,$
$21.53,$ and $23.28$ \AA. For each $z_{m},$ five independent
runs were carried out with different distributions of initial atomic velocities. As seen in this figure, the force displacement curves show very similar behavior on loading but exhibit significant differences from one run to the next at the same $z_m.$  This can be seen from the different elongations of the tensile parts of the
force-displacement curves in Fig.~\ref{fig:Spaghetti-curve}. Some of these curves show
a long tail from the minimum to the point of separation (this is what characterizes a
material as ductile-like here). Other curves with the same $z_m$ show an abrupt separation (a sharp approach to zero force). For
those cases we refer to the system as brittle-like, since the sharp drop in the tensile force to zero is related to an abrupt separation of the
asperity from one of the substrates. 


%

\section{Contact statistics}

In order to compare the different properties for the same asperity contact system, we assign each run  with the same $z_m$ a unique identifier $1\le i \le 5$.
We focus on the statistical properties of the contact dynamics of the Ru system 
by analyzing simulation results for five independent runs for eight different $z_m$.
First, we describe the $z_{m}=23.28\textrm{\AA}$ case in detail.
This case provides a representative summary of the overall qualitative features of the behavior for all $z_m$ studied in this work.

\begin{figure}
\includegraphics[clip,scale=0.32]{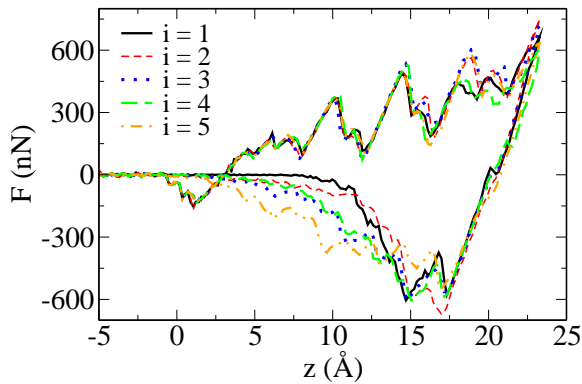}
\caption{Five independent runs for the case in which $z_{m}=23.28\textrm{\AA}.$ The double dotted-dashed line (run $i=5$) has a long tail, characterizing a ductile-like behavior. In contrast, the bold solid line (run $i=1$) sharply approaches the zero-force level, indicating a brittle-like crack, Other curves indicated by the bold dashed line (run $i=4$), thin dashed line (run $i=2$),
and dotted line (run $i=3$) correspond to various intermediate behavior.
See the text for more details.
\label{fig:Five-cases-for}}
\end{figure}

Figure \ref{fig:Five-cases-for} shows our results for $z_{m}=23.28\textrm{\AA}$
for each of five runs, each one having different initial atomic velocities. This figure shows that
during the loading stage, the curves are insensitive to the
initial atomic conditions of the system. After
the maximum load is reached and the sign of the velocity of the upper substrate reverses, 
all curves experience a near linear drop, reaching a local minimum (maximum tensile force at  $z\approx17\textrm{\AA}$).  Beyond this point, however, the curves begin showing considerable deviation from one another. The sharp drop of the tensile force represented by 
the bold solid line ($i=1$) is characteristic of brittle-like systems. On the other hand the doubled dot-dashed
line ($i=5$), the tensile force shows a long tail indicating significant necking (plastic stretching), which is a signature of a ductile-like
material. The other three curves (bold dashed, thin dashed, and dotted, $i=2-4$) correspond to intermediate cases between a brittle-like and a ductile-like behavior.

\begin{figure*}

\includegraphics[clip,scale=0.25,angle=-90]{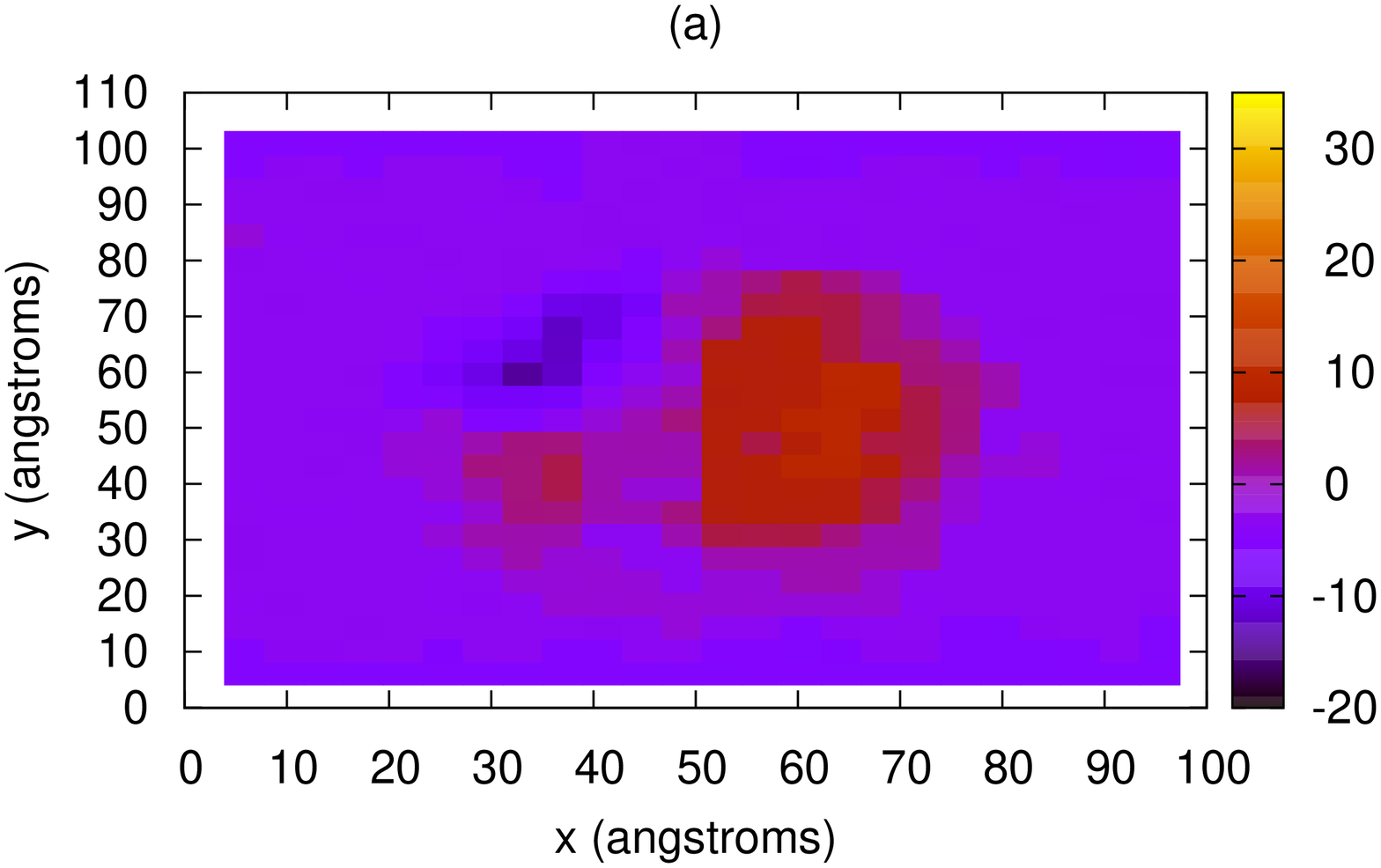}
\includegraphics[clip,scale=0.25,angle=-90]{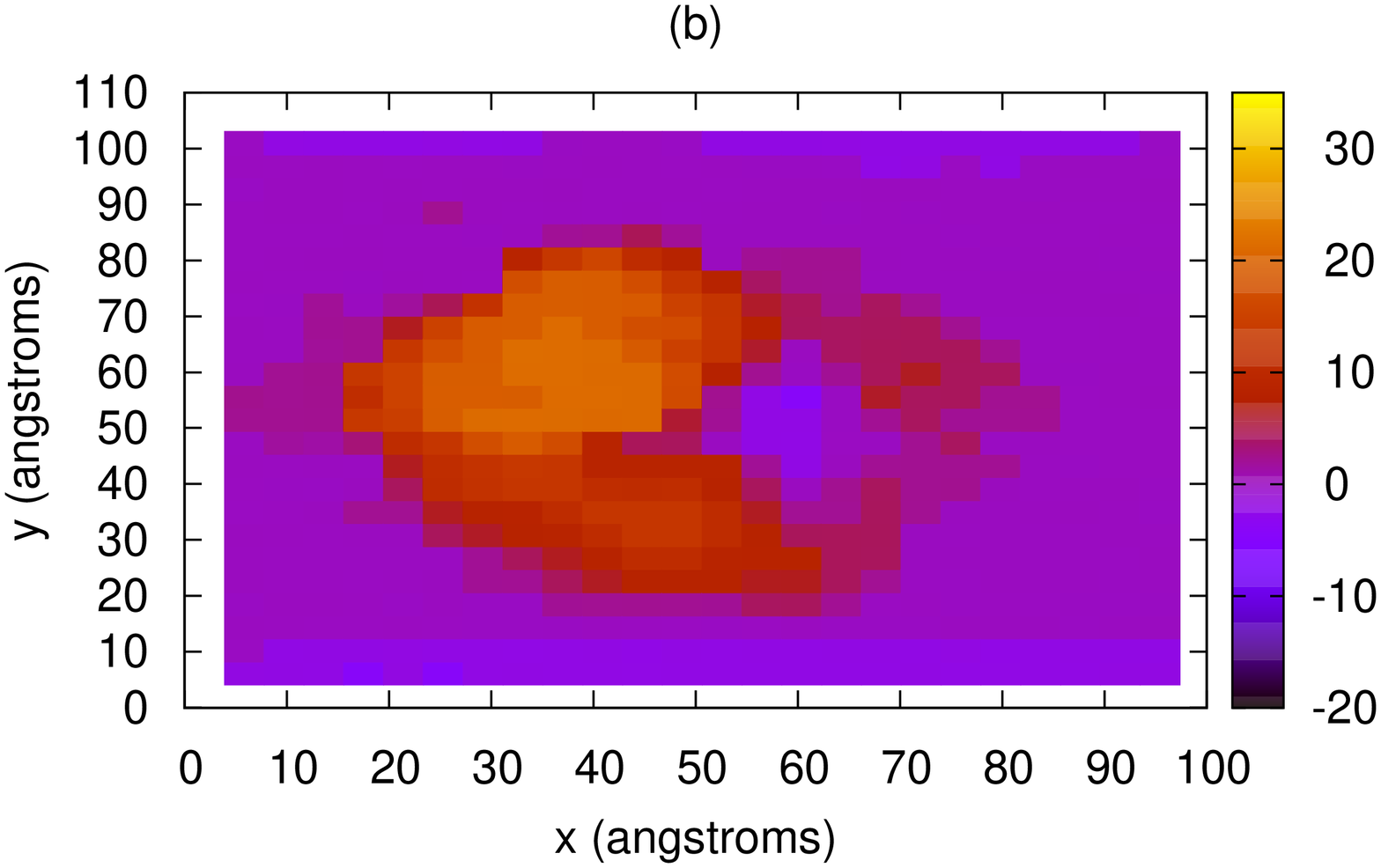}

\includegraphics[clip,scale=0.25,angle=-90]{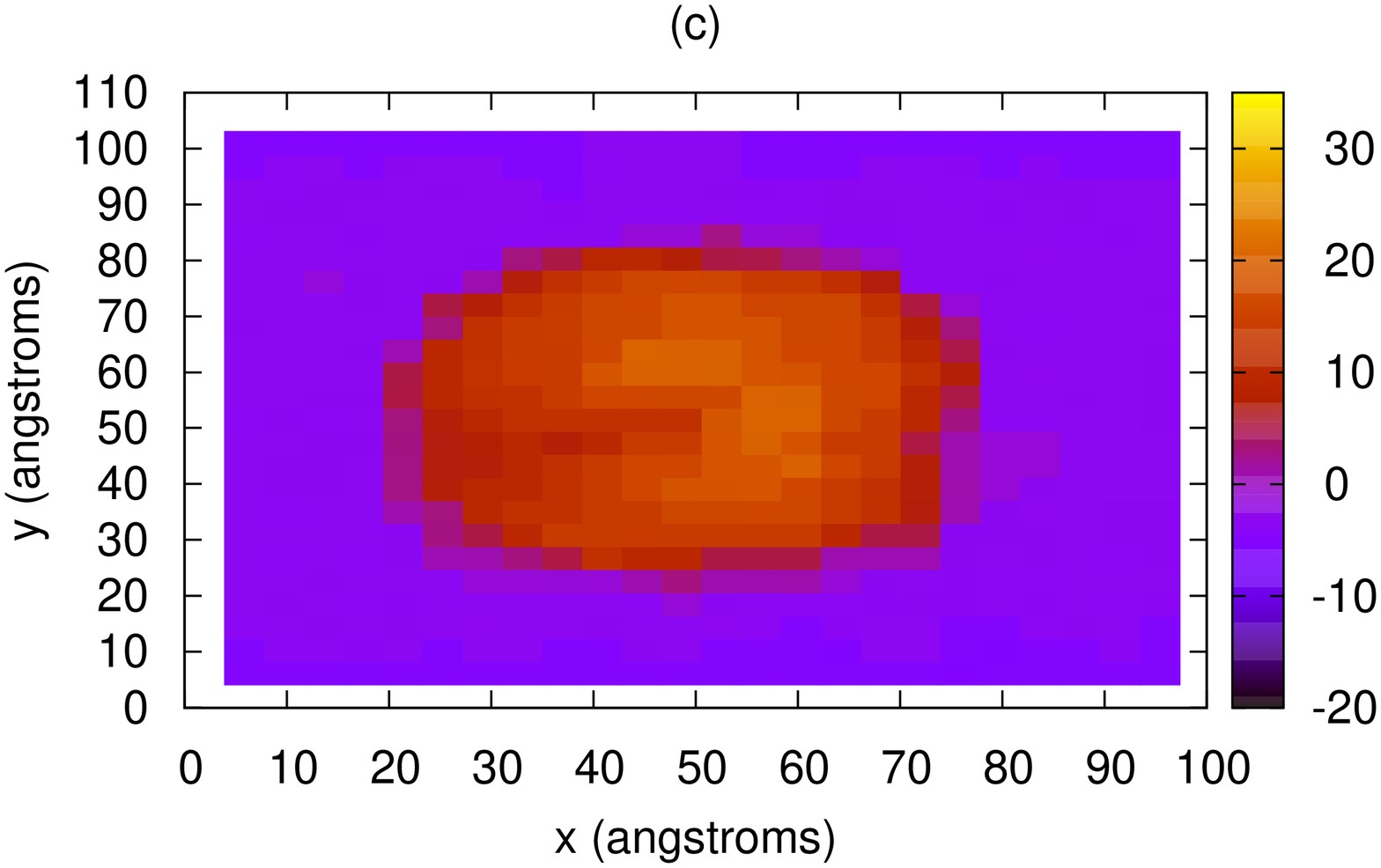}
\includegraphics[clip,scale=0.25,angle=-90]{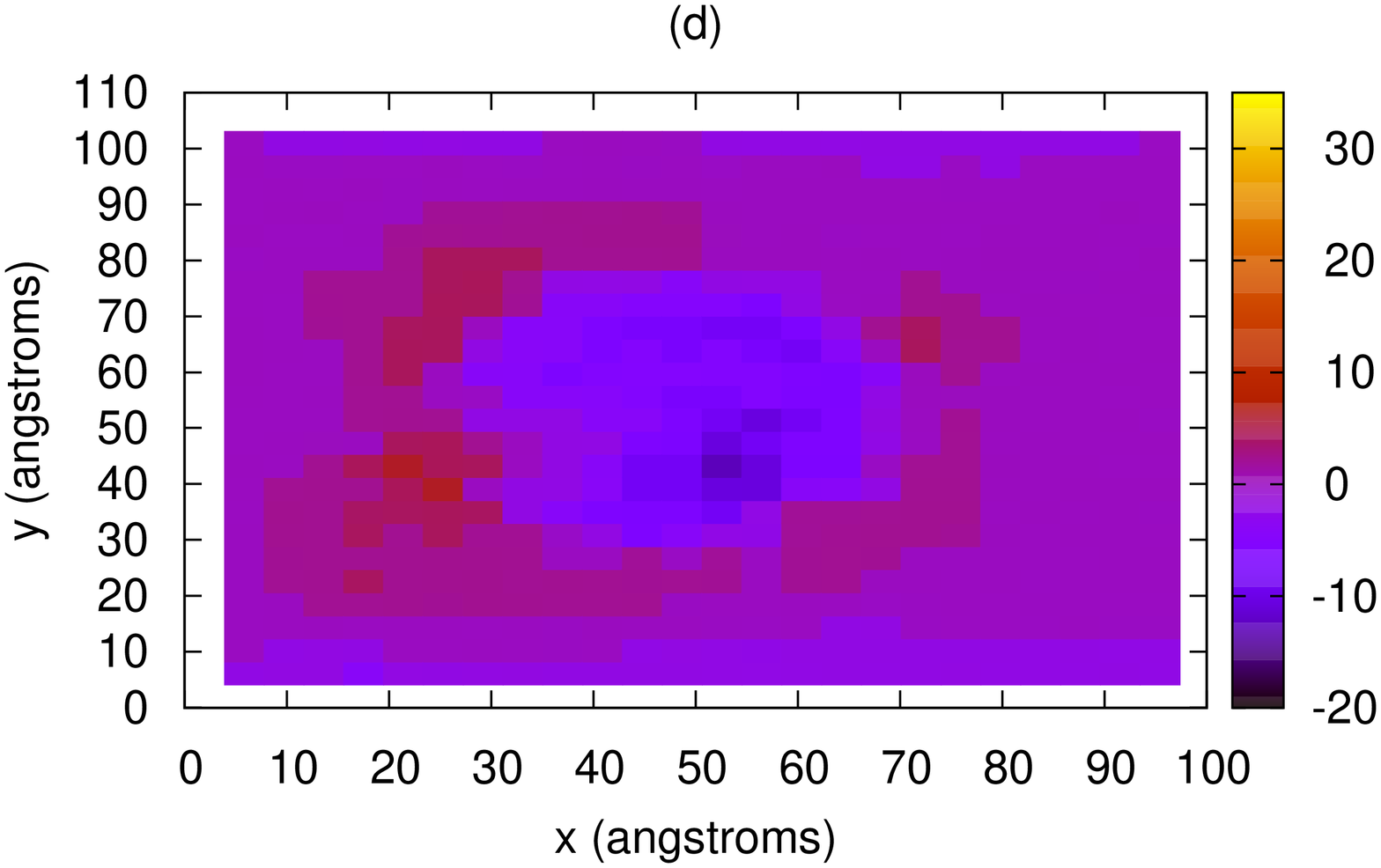}

\caption{ Topographic maps of the opposing surfaces after complete separation for $z_m=23.28\textrm{\AA}$.  
Panels (a) and (b) illustrate the case of ductile-like separation corresponding to the doubled dot-dashed $i=5$ line in Fig.~\ref{fig:Five-cases-for}. Panels (c) and (d) illustrate the case of brittle-like separation corresponding to the bold solid $i=1$ line 
in Fig.~\ref{fig:Five-cases-for}. Panels (a) and (c) represent the final shape of the upper surfaces in Fig.~\ref{fig:brittle-like-ductile-like}), while panels (b) and (d) represent the final shape of the lower surface (where the asperity is at the beginning of the simulation). The color bars indicate the height of the surface (in \AA), measured from the average surface height. Red and yellow regions correspond to parts of the surfaces protruding away from the bulk of the material, while dark blue colors represents  depressions in the surface toward the bulk of the material. 
\label{f:contacts}}
\end{figure*}

It is important to characterize the surfaces after complete separation of the two substrates,  in 
order to understand the morphological consequences of  brittle-like and ductile-like behavior, as indicated in the force-displacement curves. Changes in surface morphology may have important consequences for the behavior of high frequency MEMS/NEMS switches which are subjected to repetitive contact formation and separation. Figure~\ref{f:contacts} show topographic maps (surface heights a function of $x$ and $y$) of the two opposite surfaces in the most ductile-like and brittle-like cases seen in Fig.~\ref{fig:Five-cases-for} (i.e., bold dot-dashed $i=5$ and solid $i=1$ curves). In the ductile-like case, both surfaces [Figs. \ref{f:contacts} (a) and (b)] have large asperities with peaks in different positions. This final shape suggests that the metallic bridge between the opposite surfaces shears apart. In contrast, after brittle-like separation we see a large asperity on one substrate opposing a shallow indentation on the other substrate [Figs. \ref{f:contacts} (c) and (d)] . This final shape suggests formation of a crack near the contact between the asperity and one of the two surfaces. This crack may propagate through the pyramidal defects which form during the compression below the base of the initial asperity or deep inside the opposite surface deformed by the approaching asperity. Such defects were observed in Ruthenium contacts in Ref.~\cite{Fortini:2008}. Interestingly, crack propagation may lead either to the asperity being retained on the substrate from which it originally came or be transferred to the opposing substrate surface, depending upon which side of the initial asperity the crack formed. If the crack goes below the base of the initial asperity as in Fig. \ref{f:contacts}(d), the entire asperity is transferred to the opposite side. In contrast, in other cases, the crack was observed to pass near the intersection of the asperity and the opposing surface. In this case, little material is transferred between the opposing surfaces.           

\begin{figure}

\includegraphics[clip,scale=0.35]{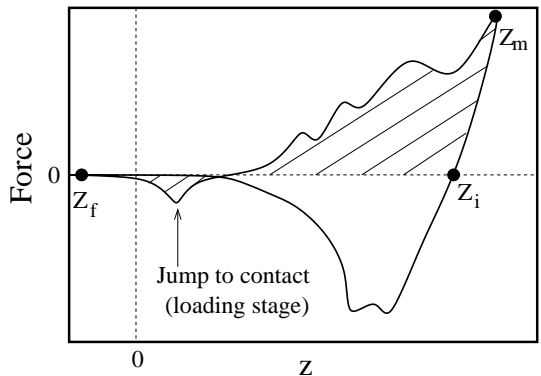}

\caption{ Schematic representation of a typical force versus displacement curve in the loading-unloading 
process. The area of the hachured region is numerically equals to the work of compression whereas the area between
the $F(z)$ curve and $F = 0$ line stands for the work of separation.  $z_i$ is the first 
point in which the force becomes zero after the maximum compression ($z_m$) and $z_f$ 
is the point in which $F = 0$ again. See the text and equations (\ref{eq:Work-separation}) and (\ref{eq:Work-compression})
for details.
\label{fig:work}}
\end{figure}

\begin{figure*}
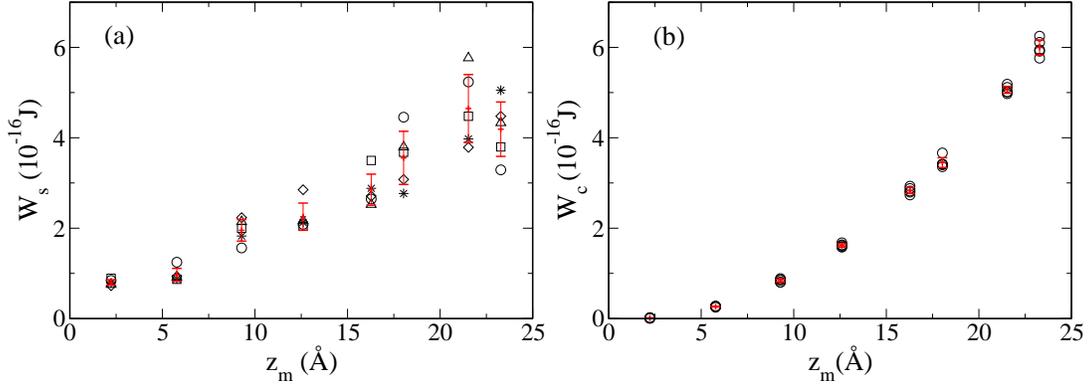

\includegraphics[clip,scale=0.3]{Work-Sep}\includegraphics[clip,scale=0.3]{Work-Compress}

\caption{Work of (a) separation and (b) compression versus maximum displacement, $z_m$. The work of
separation is the work expended to separate the two surfaces from one another, while the work of compression is the energy expended in compressing the system and then removing the applied force. The different symbols in (a) denote five simulation runs with different initial velocity distributions. In order to compare different properties of the same run, we assign each run (with the same $z_m$) a unique number $i$ from one to five. The symbols $\circ$, $\Box$, $\diamond$, $\triangle$, and $\ast$ correspond to $i=1,2,3,4,5$, respectively.
$W_s$ ($W_c$) is numerically equal to the area between $F(z)$ and the line $F=0$
in the unloading (loading) part of the force-displacement curve.  The error bars represent the standard deviation of the data shown for each $z_m$.
\label{fig:Work-of-separation}}

\end{figure*}

In order to quantify ductile-like and brittle-like behavior, we calculate the
work of separation $W_s$, which graphically corresponds to the area between
the separation curve (negative $F$) and the $F=0$ line. Mathematically this can be
written as
\begin{equation}
W_s=\int_{z_{i}(F=0)}^{z_{f}(F=0)}Fdz^{\prime}(\mathrm{unloading}), 
\label{eq:Work-separation}\end{equation}
where (unloading) indicates negative $dz$ while $z_i(F=0)$ indicates the
point on the force-displacement at which $F$ becomes for the first time equal
to zero during the unloading as indicated on Fig. \ref{fig:work} and $z_f(F=0)$ indicates
the displacement at the end of simulation at when the surfaces are completely separated and hence $F=0$. Note that $W_s>0$, since during unloading $F(z)<0$ for any $z$ between $z_f(F=0)$ and $z_i(F=0)$ and the lower limit of integration, $z_i(F=0)$, is lager than the upper limit of integration, 
$z_f(F=0)$.
We also compute the work of compression or loading
\begin{equation}
W_c=\int_{0}^{z_m}Fdz^{\prime}(\mathrm{loading})+ \int_{z_m}^{z_{i}(F=0)}Fdz^{\prime}(\mathrm{unloading}), 
\label{eq:Work-compression}\end{equation}
where (loading) indicates positive $dz$.
Since $z_m >z_i(F=0)$ but $F(z)>0$, the second integral in (3) is negative.
The results of the work of separation and the work of compression versus the maximum displacement
$z_{m}$ are shown in Fig.~\ref{fig:Work-of-separation}. From this
figure, we see that, in general, the amount of energy necessary to separate
the two surfaces increases with the maximum displacement $z_m$. However, the work of separation shows
large fluctuations, corresponding to the difference in energy required to separate the surfaces in a brittle-like (lower $W_s$) or ductile-like (higher $W_s$) manners. Comparing $W_s$ for various runs with $z_m=23.28\textrm{\AA}$ with the force displacement curves (Fig.~\ref{fig:Five-cases-for}), we see that indeed the curve with the longest tail ($i=5$) corresponds to the maximal $W_s$, while the curve with a rapid fall in the tensile force to zero ($i=1$) corresponds to the minimal $W_s$. Other curves correspond to intermediate values of $W_s$. Comparing $W_s$ for other $z_m$ with the corresponding force-displacement curves, we find that the largest values of $W_s$ correspond to the longest tails in the force-displacement curve, 
while the smallest values of $W_s$ correspond to the most abrupt decrease of the tensile force. Thus $W_s$ is a good overall indicator of the ductile-like versus brittle-like behavior. This is consistent with the definition of ductile-like versus brittle-like behavior in bulk materials.

\begin{figure*}
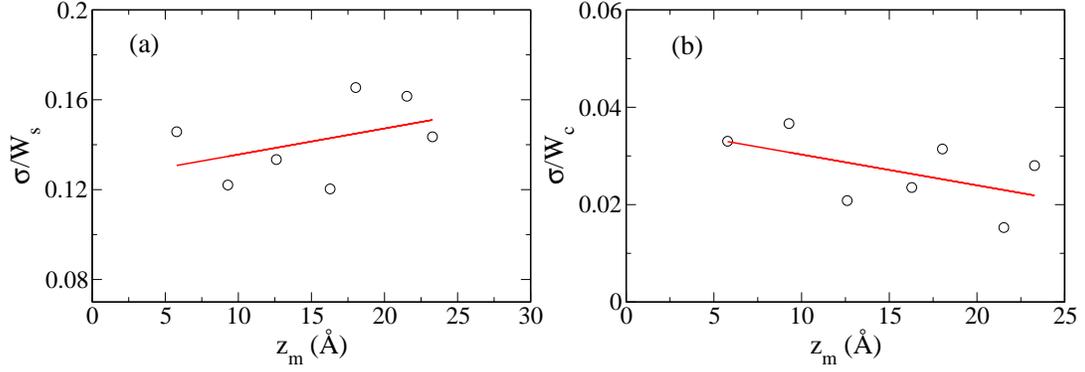

\includegraphics[clip,scale=0.3]{SigOverWorkSep.eps}\includegraphics[clip,scale=0.3]{SigOverWorkComp.eps}

\caption{ The standard deviation, $\sigma,$ of the data in Fig.~\ref{fig:Work-of-separation} divided by the work of
(a) separation and (b) compression, versus $z_{\mathrm{m}}$. The quantity 
$\sigma/W$ is the relative deviation of the work over five different runs for the same $z_m$. This data shows
that $\sigma$ is proportional to $W_s$ and $W_c$ for different $z_{m}$.  The straight lines in the figures represent linear regression results.  In both cases, the slopes are very small;   $1.16\times10^{-3} \AA^{-1}$ and $-6.32\times10^{-4} \AA^{-1}$ for panels (a) and (b), respectively. 
The data for $z_{m}=2.23\textrm{\AA}$ were not plotted since the total work done was very small (nearly elastic deformation).
 \label{fig:sigma-over-W}}

\end{figure*}

The error bars in Fig.~\ref{fig:Work-of-separation} represent the standard deviation
of the $W_s$ data over the five runs for each maximum displacement: 
$\sigma=\left[\left(1/N\right)\sum_{i=1}^{N}\left(W_{s,i}-\overline{W}_s\right)^{2}\right]^{1/2},$
where $\overline{W}_s=(1/N)\sum_{i=1}^{N}W_{s,i}$ is the average
over $N=5$ independent measurements of $W.$  As the degree of
compression ($z_m$) increases, the magnitude of the error bars grow. On the other hand, the relative values of $\sigma$
(i.e., $\sigma/W_s$) are nearly independent of $z_{m},$ as seen from Fig. \ref{fig:sigma-over-W}(a).
Figure \ref{fig:sigma-over-W}(b) shows an analogous graph for the work of compression, which indicates that the relative fluctuation of the work of compression is only approximately a quarter of that for the relative fluctuation of the work of separation. Linear regression yields negligible regression coefficients in the work of separation and compression cases,
$1.16\times10^{-3} \AA^{-1}$  and $-6.32\times10^{-4} \AA^{-1}$, respectively (see the lines in Fig. \ref{fig:sigma-over-W}).    

\begin{figure}
\includegraphics[clip,scale=0.32]{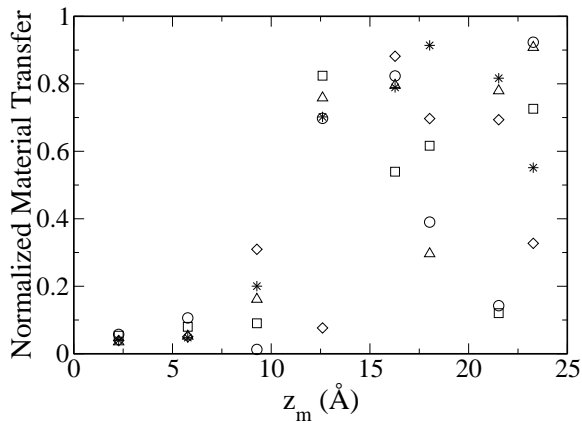}

\caption{Normalized number of asperity atoms which remain attached to the upper substrate after separation (the number of atoms transferred to the upper substrate divided by the total number of atoms in the asperity, $N=2981$)
versus $z_m$. See the text for more details.
\label{fig:mat-transf}}
\end{figure}

We also analyze the number of asperity atoms which are transferred from the lower to upper substrate upon separation - see Fig. \ref{fig:mat-transf}.  For small degrees of compression, only a small portion of the asperity is transferred to the upper substrate. This is because at very low compression, the deformation of the system is nearly elastic and the contact area is small. As the degree of compression increases, more asperity atoms become in contact with the upper substrate, increasing the degree of adhesion and contact area between the two surfaces -- resulting in larger material transfer. We also observe that the fluctuation in the number of asperity atoms transferred to the upper substrate varies greatly, especially at large $z_{m}$ ($>12.6\textrm{\AA}$), going from $10\%$ to $90\%$ of the asperity transferred in some cases.  

\begin{figure}
\includegraphics[clip,scale=0.32]{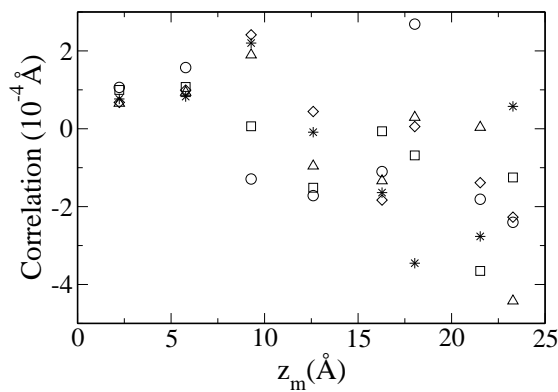}

\caption{Correlation between the stalagmites and stalactites heights on opposing surfaces
versus the maximum displacement in the $z$-direction. See the
text for details.
\label{fig:corr}}
\end{figure}

Stalactites and stalagmites are formed on the opposing surfaces during the separation process. The heights of these structures vary dramatically from run to run (at the same $z_m$). One way to characterize the topography of the surfaces after separation is to compute the cross-correlation between the heights at the same $(x,y)$ coordinates of these landscapes. In order to do this, we divided the system into a $25 \times 25$ grid in the $xy$ plane, in which each grid cell is a rectangle of dimensions $\Delta x=L_x/25 $ and $\Delta y=L_y/25$.  For each cell we find the atom with the minimal $z=-h_u(x,y)$ on the upper substrate and  the atom with maximal $z=h_l(x,y)$ in the lower substrate [where the surface profiles are $h(x,y)$]. Then the cross-correlation is computed as
\begin{align*}
\Psi=\frac{1}{N_{\Omega}}\sum_{\{x,y\in\Omega\}}h_{u}(x,y)h_{l}(x,y)\\
-\frac{1}{N_{\Omega}^{2}}\sum_{\{x,y\in\Omega\}}h_{u}(x,y)\sum_{\{x,y\in\Omega\}}h_{l}(x,y),
\end{align*}
where $\Omega$ is the set of cells which belong to the contact area at the point of maximal compression, 
and $N_\Omega$ is the number of such cells. $N_\Omega$ typically goes from 136 to 225, depending 
on the magnitude of $z_m$.  The area $N_\Omega\Delta x\Delta y$ has the physical meaning of maximum contact area. 
We implement this condition in order to compute correlations only for the area directly affected by the contact. 
Our cross-correlation definition is devised in a such a way that the typical shape observed in the ductile-like samples 
[Fig.~\ref{f:contacts}(a),(b)], in which two asperities face each other,
yields positive correlation, while the final shape resulting from brittle-like behavior
[Fig.~\ref{f:contacts}(c),(d)], in which an asperity on one side faces a depression on the other, yields negative correlation. Comparing Fig.~\ref{fig:corr} and  Fig.~\ref{fig:Work-of-separation} we see that the
runs with the largest positive correlations always coincide with the runs with
the largest work of separation. This observation is consistent with the expectation that ductile-like behavior is associated with the formation of metallic bridges (between the opposing surfaces) which elongate as the substrates are pulled apart. When the bridge finally breaks, the stalactite and the stalagmite formed on the opposite surfaces facing one another. Brittle behavior is associated with crack formation which commonly propagates through the defects formed below the surface during compression, resulting in the asperity on one side opposing a depression on the other.

%
%


\section{Discussion}

\subsection{Effect of Initial Conditions}

The main question we address in this article is how the thermal noise alters the
behavior of nanoasperity contacts. 
Clearly, large structural changes in the initial asperity can lead to much greater variability than thermal noise. 
An example of such a study was given in Ref.~\cite{Song2007} where the force displacement curves for multiple loading-unloading cycles of Au nanoasperity contacts were calculated for different starting nanoasperities.
The force displacement curves were found to be drastically different for different nanoasperities, with the maximal tensile force varying by more than factor of two. 

The changes in initial conditions due to thermal noise are much more subtle but as shown in the previous section they have a strong effect during the separation stage.
In order to address the question of how the initial conditions affect the results in the loading-unloading stage in greater detail, we  perform five additional simulations in which we assign different random velocities to the atoms before the annealing of the asperity resulting in five different initial asperities. 
We than assign to the atoms in each asperity the same random velocities and perform the loading-unloading cycle. 

We also perform five additional runs starting from five different asperities annealed for 200$ps$ instead of the standard annealing time of 100$ps$ as describe in section~\ref{sec:met}.
A longer annealing time guaranties an even larger structural variability. Each asperity is annealed starting from different random velocities, while the same random velocities were assigned to all asperities before the loading-unloading cycle. 

\begin{figure}
\includegraphics[clip,scale=0.32]{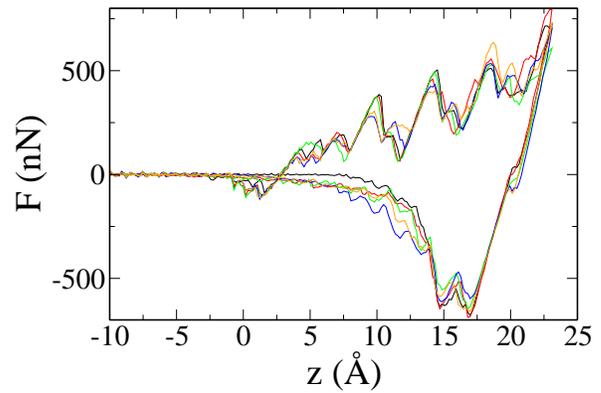}

\caption{Effect of the assigning different random velocities to the atoms prior to the annealing of the asperities so that the resulting asperities have slightly different shape.  We plot force over the asperity atoms against displacement in the $z$-direction. Each curve corresponds to a different asperity prepared with the standard annealing time.\label{fig:normal} We do not see any qualitative differences
with the previously studied case in which the simulations differ only by the
atomic velocities at the start of the compression stage~[Fig.~\ref{fig:Five-cases-for}]}

\end{figure}

\begin{figure}
\includegraphics[clip,scale=0.32]{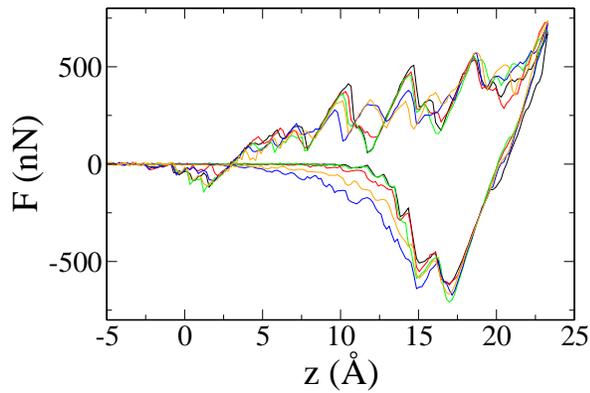}

\caption{Same as figure  \ref{fig:normal} but for the two times longer annealing procedure. Force over the asperity atoms against displacement in the $z$-direction.
Each curve corresponds to a different asperity prepared with the longer annealing time of 200$ps$.
\label{fig:2xlonger}}

\end{figure}

The results for the standard annealing can be seen in Fig. \ref{fig:normal} while the result for the asperities with  longer annealing time are shown in Fig.
\ref{fig:2xlonger}. In both cases, it is clear that the tail of
the tensile force varies in a way similar to the results shown in Fig.~\ref{fig:Spaghetti-curve}.
As  expected the variability is more evident for the asperities annealed for a longer time.

\subsection{Effect of Asperity Size}

\begin{figure}
\includegraphics[clip,scale=0.32]{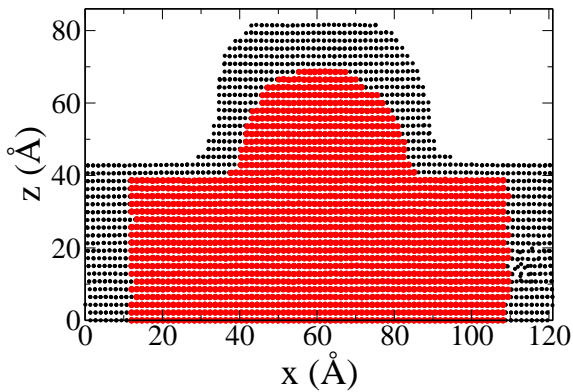}

\caption{Comparison between the large system and the small one considered in section~\ref{sec:met} 
\label{fig:different-size-systems}}

\end{figure}

We investigate the effect of asperity size on the loading-unloading curves.
We prepared a larger system consisting of
111664 atoms, with 7000 atoms in the asperity. 
The comparison between the structures of the large system and the smaller one considered in section~\ref{contacts} can be seen in Figure \ref{fig:different-size-systems}, in which the upper substrate was removed for better viewing purposes. 

\begin{figure}
\includegraphics[clip,scale=0.32]{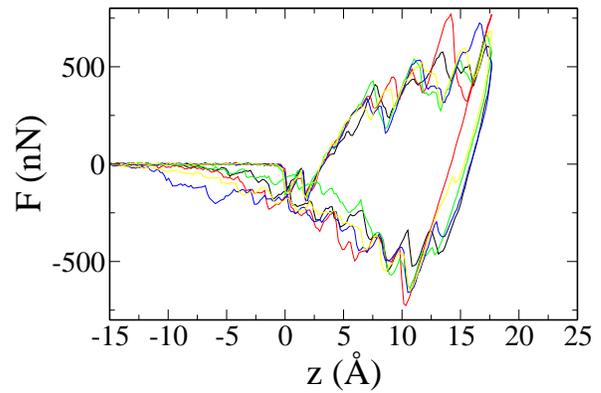}
\caption{Force over the asperity atoms against displacement in the $z$-direction
for the large system.}
\label{fig:190-big}
\end{figure}

\begin{figure}
\includegraphics[clip,scale=0.32]{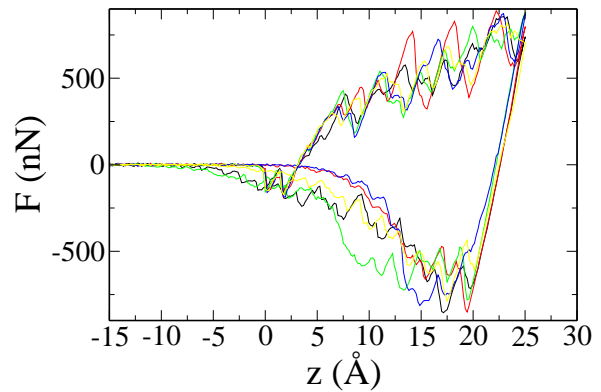}

\caption{The same as in Fig. \ref{fig:190-big} but for maximal compression,
 25.0 \AA. }
\label{fig:232-big}
\end{figure}

Five big asperities were prepared according to the standard annealing procedure described in section~\ref{sec:met} and with five different initial random velocities. The same initial velocity distribution
was used for the five asperities in the loading-unloading cycle. 
Here, only two maximal compressions were considered:
17.6 \AA~ and 25.0 \AA. The results are shown in Fig.~\ref{fig:190-big} and Fig.~\ref{fig:232-big} respectively. 
We find a high degree of variability, independently of system size and degree of compression.

\subsection{Gold statistics}
\begin{figure}
\includegraphics[scale=0.45]{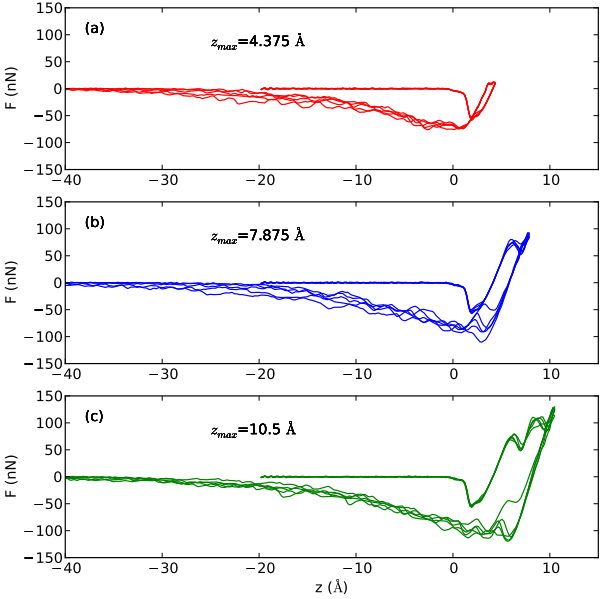}

\caption{Force versus displacement for Au contacts.
Shown are 5 independent runs for three different maximum displacements $z_{max}$.
\label{fig:au}}

\end{figure}

\begin{figure}
\includegraphics[width=8cm]{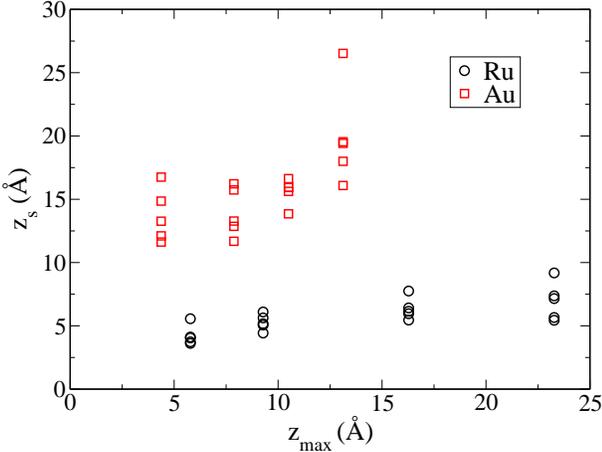}

\caption{Effective tensile strains $z_s=W_s/|F_{mt}|$ for five runs and different maximal compressions $z_{\mathrm{max}}$  for Ru and Au, where $|F_{mt}|$ is the maximal tensile force achieved in a specific run. \label{fig:comparison-ru-au}}

\end{figure}

Due to its popularity as a contact material, and as a complement to our previous work~\cite{Fortini:2008} we performed simulation to understand the contact statistic of Au.
We performed simulations of Au contact loading and unloading for different starting atomic velocities using the Au asperity generated as described in Ref.~\cite{Fortini:2008}. Fig.~\ref{fig:au} shows the force-displacement curves for three different maximum displacement. For each displacement we produced five runs assigning different initial velocities to the atoms before the loading-unloading stage.
Like in Ru we do not find any variability in the loading phase. During unloading, a degree of variability is present, but most notably there is no change of the mechanism of separation. That is, in all cases we observe a ductile-like transition.
To compare these results with those of Ru, we calculate the effective tensile strains $z_s=W_s/|F_{mt}|$ for five runs for Ru and Au, where $W_s$  is defined in Eq. (\ref{eq:Work-separation}) and $|F_{mt}|$ is the maximal tensile force achieved in a specific run.  Figure~\ref{fig:comparison-ru-au}
shows different values of $z_s$ for different maximal displacements for both materials. 
We see that the values of $z_s$ for Au are typically  three times larger than for Ru which emphasizes 
the much higher ductility of Au with respect to Ru. 
On the other hand the relative variability of $z_s$ are almost the 
same for Ru and Au. For each $z_m$ we find $\sigma_z$, the standard deviation 
of $z_s$, and compare the relative standard deviation $\sigma_z/z_s$ for
Au and Ru averaged over four different $z_m$. We see that this quantity for
Au is 0.14 while for Ru it is 0.17. Thus the presence of two different modes
of separation for Ru contacts does not make them more variable than Au contacts.

\section{Conclusion}

We find substantial variability in the way two ruthenium surfaces separate after contact.  
This variability is a reflection of the different modes of separation available to the system (ductile-like versus brittle-like) and of the fact that these different modes are competing with one another. For the same degree of compression, simulations (that only differ in the initial thermal velocities of the atoms) can lead to either brittle-like or ductile-like behavior, as seen in Figs.  \ref{fig:brittle-like-ductile-like}, \ref{f:contacts}, \ref{fig:Work-of-separation}, and \ref{fig:corr}. The ductile-like behavior can be described as the formation of a bridge between the contacting substrates that narrows via plastic deformation and eventually separates via a shear process.  This leads to the formation of asperities on both facing surfaces.  The  brittle-like behavior may be characterized as the propagation of a crack through the substrate in a region that is damaged by plastic deformation, resulting in an asperity on one surface facing a depression on the opposite surface.  The work of separation is much greater for the case of ductile-like separation than for the case of brittle-like separation.  While the work of separation (as well as its fluctuations) gradually increases with the degree of compression $z_m$, the surface cross-correlation,  as well as material transfer shows a dramatic increase of fluctuations for large compression (i.e., $z_m>12\AA$ or $\sim3$ atomic layers). Below this value, the material transfer is always small(constituting a small part of the initial asperity), while above this threshold the magnitude of the material transfer becomes highly variable, ranging from 10\% to 90\% of the original asperity.  Morphology cross-correlation analysis shows that in the large compression case, this variability can be attributed to the difference in separation behavior - ductile-like versus brittle-like.  The ductile-like contact separation mode seen in ruthenium is akin to the separation behavior seen in gold.\cite{Fortini:2008} On the other hand, the brittle-like separation mode observed here is uncommon in ductile materials.  This makes ruthenium behavior very distinct from the behavior of gold. Further, contacts made from materials that exhibit predominantly brittle-like behavior will tend to require lower work of separation than those made from ductile-like contact materials.  Some plastic deformation is, however, desirable, since good electrical contact performance requires large contact area and the availability of some plasticity will lead to an increase in contact area. In conclusion,
although Ru contacts have two modes of separations, brittle-like and ductile-like they are more reliable than Au contacts because the effective separation
distance $z_s$ for Ru contacts is several times smaller than for Au contacts and
the variability of $z_s$ and $W_s$ for Ru and Au are almost equal. Large values
of $z_s$ for Au indicate formation of large stalactites and stalagmites 
on the opposite contact surfaces after separation, leading to the increasing roughness of the surfaces after multiple switching cycles \cite{Song2007} and finally to large 
fluctuations of the stiction force and fast wearing of the contacts.
In contrast, in Ru the asperities remain relatively small and although some 
degree of uncertainty is present due to two possible modes of separation, 
this uncertainty is comparable with that of Au contacts, 
which exhibit only ductile-like behavior.



\section{Acknowledgments}
The authors thank Dr. Jun Song for useful discussions. A.B.O. thanks the Brazilian science agencies CNPq and FAPEMIG for financial support.
S.V.B. thanks the Office of Academic Affairs of Yeshiva University for funding the Yeshiva University high performance computer cluster and acknowledges the partial support of this research through the Dr. Bernard W. Gamson Computational Science Center at Yeshiva College.  A.F thanks the DFG for support via SFB840/A3.
This work was supported by the UCSD/NEU DARPA S$\&$T Center for  MEMS Reliability and Design Fundamentals Grant No. HR0011-06-1-0051 and Dr. Dennis Polla, program monitor.

\bibliographystyle{aip}
\bibliography{contact}

\end{document}